\documentclass[twocolumn]{aastex701}
\usepackage{amsmath}
\usepackage{svg}
\usepackage{nicefrac}
\usepackage[none]{hyphenat}
\journalinfo{}

\begin{document}

\title{Using GALEX UV Excess to Search for Metal-poor Halo Stars}

\author[0009-0000-8325-9736]{Chase L. Smith}
\affiliation{Department of Physics and Astronomy, University~of~Wyoming, 1000~E.~University~Ave., Dept.~3905, Laramie, WY 82071, USA}
\email{csmit151@uwyo.edu}

\author[0000-0002-0870-6388]{Maxwell~Moe}
\affiliation{Department of Physics and Astronomy, University~of~Wyoming, 1000~E.~University~Ave., Dept.~3905, Laramie, WY 82071, USA}
\email{mmoe2@uwyo.edu}

\author[0009-0002-2866-9788]{Megan Frank}
\affiliation{Department of Physics and Astronomy, University~of~Wyoming, 1000~E.~University~Ave., Dept.~3905, Laramie, WY 82071, USA}
\email{mfrank13@uwyo.edu}

\author[0009-0009-0960-6280]{Raven Cilley}
\affiliation{Department of Astronomy, University of Michigan, 1085 S.~University Ave., Ann Arbor, MI 48109, USA}
\email{rcilley@umich.edu}

\author[0009-0002-5503-3670]{Javier Fregoso}
\affiliation{Applied Physics Program, CSU Channel Islands, 1 University Dr., Camarillo, CA 93012, USA} 
\email{javier.fregoso730@myci.csuci.edu}

\author[0009-0003-3096-7618]{Alexander Gleason}
\affiliation{Homer L. Dodge Department of Physics and Astronomy, University of Oklahoma, 440 W. Brooks St., Norman, OK 73019, USA}
\email{}

\author[0009-0005-4279-4093]{Ella Morton}
\affiliation{Department of Physics, Reed College, 3203 SE Woodstock Blvd, Portland, OR 97202, USA} 
\email{ellamorton@reed.edu}

\author[0009-0009-9916-792X]{Grace Nelson}
\affiliation{Department of Physics, Michigan Technological University, 1400 Townsend Dr., Houghton, MI 49931, USA}
\email{genelson318@gmail.com}

\author[0009-0005-0016-0463]{Mary~Kate Petrykovets}
\affiliation{Department of Astronomy, San Diego State University, 5500 Campanile Drive, San Diego, CA 92182, USA}
\email{mpetrykovets1872@sdsu.edu}

\author[0009-0007-2285-9528]{Daniel Reshan}
\affiliation{Department of Physics, Case Western Reserve University, 10900 Euclid Ave, Cleveland, OH 44106, USA}
\email{d.reshan626@gmail.com}

\author[0009-0005-6026-765X]{Kaitlyn Schultz}
\affiliation{Department of Physics and Astronomy, University~of~Wyoming, 1000~E.~University~Ave., Dept.~3905, Laramie, WY 82071, USA}
\email{kaitlynschultz@earthlink.net}

\author[0000-0002-5782-9093]{Daniel~A.~Dale}
\affiliation{Department of Physics and Astronomy, University~of~Wyoming, 1000~E.~University~Ave., Dept.~3905, Laramie, WY 82071, USA}
\email{ddale@uwyo.edu}

\author[0009-0005-0582-8469]{Nikhil Patten}
\affiliation{Department of Physics and Astronomy, University~of~Wyoming, 1000~E.~University~Ave., Dept.~3905, Laramie, WY 82071, USA}
\email{npatten@uwyo.edu}

\correspondingauthor{Chase L. Smith}
\email{csmit151@uwyo.edu}

\begin{abstract}
Metal-poor solar-type stars display a significant reduction in metal-line blanketing at short wavelengths, leading to an excess of near-ultraviolet (NUV) flux compared to their metal-rich counterparts. We utilize GALEX NUV and $\it{Gaia}$ DR3 photometry along with ground-based spectroscopy to establish a correlation between NUV excess and [Fe/H]. We construct a sample of 492 solar-type (F5-G9) halo stars with NUV excess and measured metallicitices. We perform our own observations with the KOSMOS spectrograph at Apache Point Observatory's 3.5m telescope to measure the abundances of 13 halo stars, 11 of which did not have previous metallicity measurements. Our targeted 13 halo stars span $-$2.92 $<$ [Fe/H] $<$ $-$1.97 and are all $\alpha$ enhanced with [$\alpha$/Fe] = 0.05\,-\,0.73. For our full sample of 492 objects, we find an anti-correlation between NUV excess and [Fe/H] that is statistically significant at the 8$\sigma$ level. GALEX NUV excess can be used to distinguish very metal-poor (VMP) stars ([Fe/H] $<$ $-$2) from their metal-rich counterparts. However, there is significant dispersion in the relation due to NUV chromospheric variability caused by rotational effects and magnetic cycle activity. The NUV chromospheric variability inhibits our ability to reliably distinguish extremely metal-poor (EMP) stars ([Fe/H] $<$ $-$3) from VMP stars based on photometry alone.  UV spectra of EMP halo stars are needed to better calibrate their atmospheric properties and variability. 
\end{abstract}


\section{Introduction}
\label{sec:Intro}

Metallicity is an important tracer of the stellar and galactic formation history of the Milky Way, and precise chemical abundances are essential for many astrophysical sub-fields. For instance, enhanced star formation caused by past dwarf galaxy mergers can still be seen in regions of elevated metallicity \citep{Ishigaki2013,Matteucci2021}. Metallicity also plays a pivotal role in many other physical processes, affecting stellar size, exoplanet occurrence rates, and binary fraction, among others  \citep{Moe2019,Zhu2019,Xin2022}.

Over the last decade, multiple large-scale spectroscopic surveys have been instrumental in advancing our understanding of the chemical composition of the universe. The GALAH survey \citep{Buder2021} obtained high-resolution spectra of 590,000 stars and measured abundances to an accuracy of $\approx$\,0.05 dex. Similarly, the APOGEE (\negthickspace\negthickspace\citealt{Abdurrouf2022}) survey measured abundances of 650,000 stars to a precision of $\approx$\,0.10 dex. The RAVE survey \citep{Shank2022} obtained medium-resolution spectra of 450,000 halo stars, measuring metallicities to an accuracy of $\approx$\,0.2 dex. Using a machine-learning technique, metallicities of 175 million stars were recently measured to $\approx$\,0.1 dex precision based on {\it Gaia}'s low-resolution XP spectra \citep{Rene2023}.

\begin{figure*}
    \centering
    \includegraphics[trim = 0.3cm 3.3cm 0.3cm 4.6cm, width=1\linewidth]{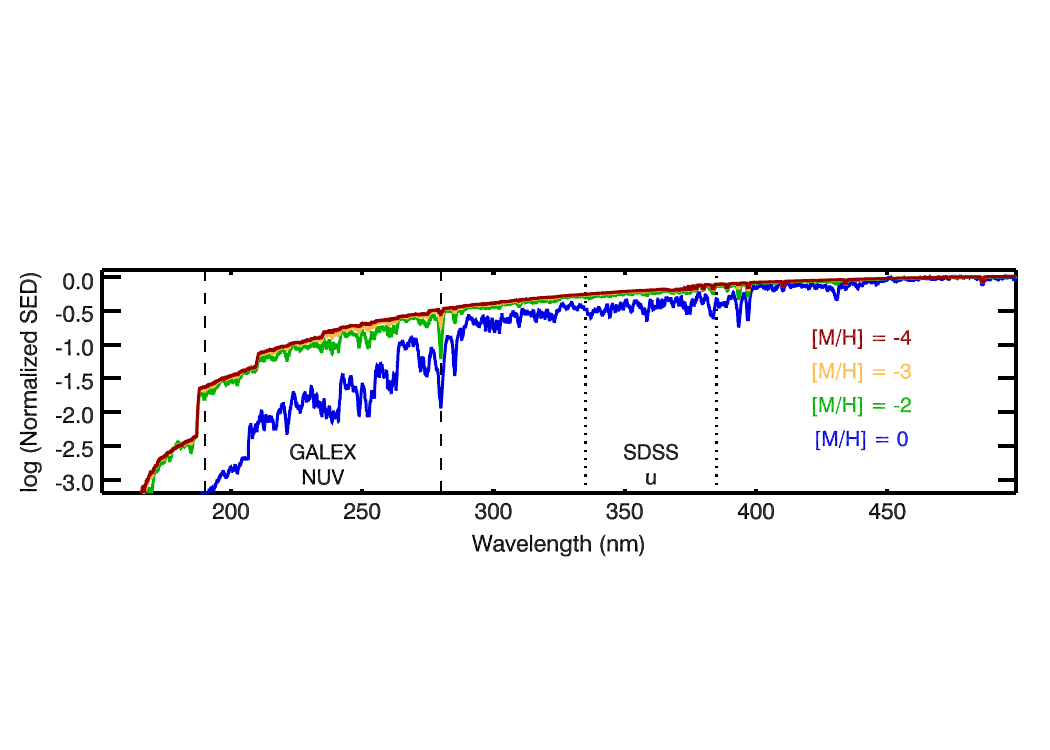}
    \caption{PHOENIX model spectra of solar-type stars at four different metallicities. Compared to SDSS u-band (dotted), the GALEX NUV passband (dashed) provides greater leverage in distinguishing VMP stars ([Fe/H]~$\le$~$-$2; green), EMP stars ([Fe/H]~$\le$~$-$3; yellow), and UMP stars ([Fe/H]~$\le$~$-$4; red).}
    \label{fig:Phoenix}
\end{figure*}

Metal-poor stars are distinguished according to their iron abundance: very metal-poor (VMP) stars have [Fe/H]\,$<$\,$-$2, extremely metal-poor (EMP) stars have  [Fe/H]\,$<$\,$-$3, and ultra metal-poor (UMP) stars have [Fe/H]\,$<$\,$-$4 \citep{Beers2005}. Only two dozen UMP stars are currently cataloged in the JINAbase \citep{Abohalima2018}. Larger samples of UMP stars are needed to test formation theories of our Milky Way. However, spectroscopic surveys are observationally expensive, especially if they target stars randomly. 

Photometric surveys can help to identify VMP, EMP, and UMP candidates. The Pristine survey \citep{Aguado2019,Sestito2020,Lucchesi2022} utilized a narrow-band Ca H and K filter with the 1-degree field-of-view imager ``MegaCam" on the Canada-France-Hawaii Telescope to identify VMP stars. Metal-poor solar-type stars display a significant reduction in metal-line blanketing, in particular iron-group line blanketing, at short wavelengths, leading to excess ultraviolet flux compared to solar-metallicity stars. \citep{Short2005}. Previous surveys in the SDSS u-band \citep{Gu2015,Ivezic2008} have been particularly successful in differentiating VMP stars from their metal-rich counterparts. Despite these photometric surveys' ability to identify metal-poor stars, they still struggle to robustly differentiate between VMP, EMP, and UMP candidates.

The Galaxy Evolution Explorer Satellite (GALEX) \citep{Martin2005} performed an all-sky imaging survey in the near-ultraviolet (NUV), thereby obtaining precise photometry for tens of millions of stars in our Milky Way. Metal-poor stars have significant UV excess within the GALEX NUV passband (180-280nm). For example, \citet{Mohammed2019} measured an ultraviolet-optical color-metallicity relation for Galactic red clump stars using GALEX and {\it Gaia}. 

To motivate the potential of GALEX NUV photometry as a tracer of metallicity, we compare in Fig.~\ref{fig:Phoenix} PHOENIX model spectra \citep{Husser2013} of solar-type dwarfs (log g = 4.5, $T_{\mathrm{eff}}$ = 5,800~K) at four different metallicities. All spectral energy distributions (SEDs) are smoothed to 10\,\AA\ and normalized at 480 nm. The solar-metallicity spectrum exhibits significant metal-line blanketing in both the SDSS u and GALEX NUV passbands. However, in the SDSS u-band, the [M/H] = $-$2 star exhibits only moderate metal-line blanketing, and the [M/H] = $-$3 and $-$4 spectra are indistinguishable. Averaged across the SDSS u-band, the [Fe/H] = $-$4 star is only 0.02 mag brighter than the [Fe/H] = $-$3 star, which in turn is only 0.06 mag brighter than the [Fe/H] = $-$2 star.  Meanwhile, in the GALEX NUV passband, the [M/H] = $-$2 star is substantially fainter and the [M/H] = $-$3 and $-$4 spectra are noticeably different. Averaged across the GALEX NUV passband, the [Fe/H] = $-$4 star is 0.07 mag brighter than the [Fe/H] = $-$3 star, which in turn is 0.24 mag brighter than the [Fe/H] = $-$2 star. Thus the GALEX NUV passband offers a greater dynamic range and superior leverage in measuring the photometric metallicities of stars. In principle, GALEX NUV photometry may potentially distinguish VMP, EMP, and UMP stars. 

The goal of this study is to measure the correlation between GALEX NUV excess and spectroscopic metallicity [Fe/H] for a population of kinematically selected metal-poor halo stars. In Section~\ref{sec:Sample}, we construct a sample of 492 halo stars with GALEX NUV photometry, {\it Gaia} data, and ground-based spectroscopic metallicities. In Section \ref{sec:Observations}, we reduce and analyze spectroscopic data for 13 halo stars that exhibit a NUV excess. We measure the abundances and physical properties of our targeted 13 halo stars in Section~\ref{sec:abundances}. In section~\ref{sec:Discussion}, we analyze the correlation between NUV excess and metallicity for our full sample. We summarize our main results in Section~\ref{sec:conclusions}.

\section{Sample Selection}
\label{sec:Sample}

\begin{figure}
    \centering
    \includegraphics[trim = 1.5cm 0.8cm 0.8cm 0cm, width=1\linewidth]{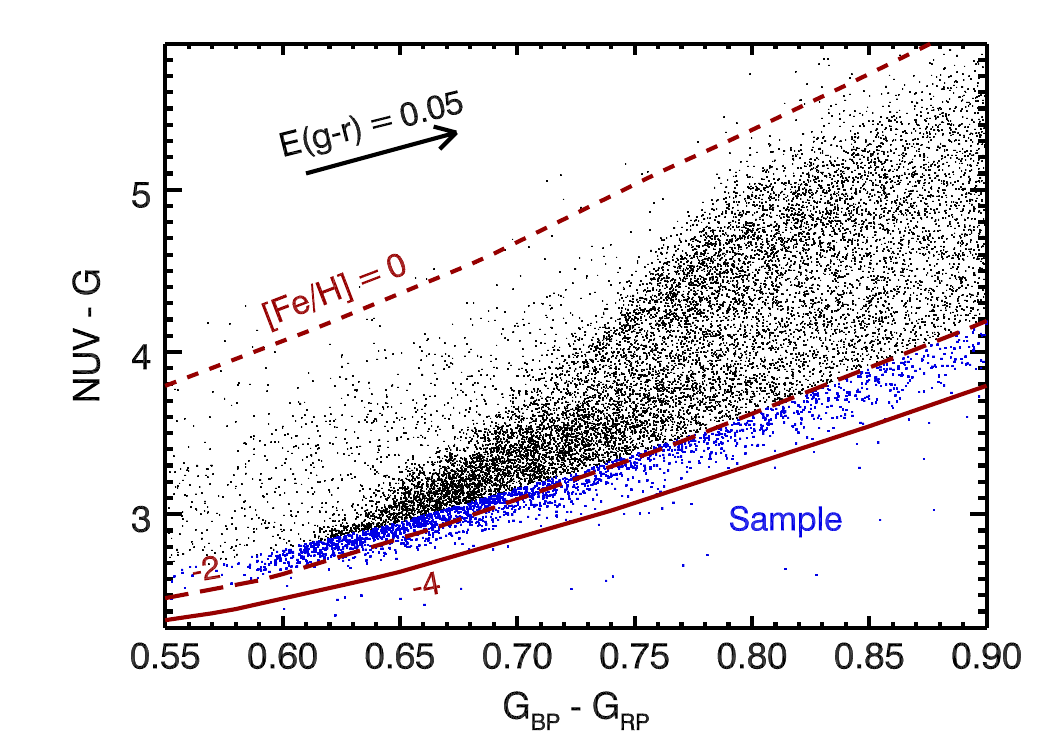}
    \caption{Color-color diagram of our selected bright halo stars with GALEX counterparts. We display MIST model colors for metallicities [Fe/H] = $-$4, $-$2, and 0 (red). A dust reddening vector of E(g-r) = 0.05 is shown with a black arrow. We initially select the 1,994 stars (blue) that exhibit a NUV excess, being either brighter than the [Fe/H] = $-$2 model and/or are within $\Delta$(NUV$-$G)$_{\rm [Fe/H]=-4}$ $\le$ 0.3 mag of the [Fe/H] = $-$4 model for the given BP$-$RP color.}
    \label{fig:HaloColorColor}
\end{figure}

We identify halo stars that exhibit a GALEX \citep{Bianchi2017,Bianchi2011} NUV excess as follows. Utilizing the {\it Gaia} archive\footnote{\url{https://gea.esac.esa.int/archive/}}, we initially select the 18,410 halo stars from the {\it Gaia} DR3 catalog (\negthickspace\negthickspace\citealt{Gaia2023}) that are brighter than G\,$=$\,14.5, are within distances $d$~$<$~2~kpc (parallaxes $\varpi$~$>$~0.5~mas), have distance measurements better than 20\% (parallax over error $\varpi$/$\sigma_{\varpi}$~$>$~5), are located at high galactic latitudes $|b|$~$>$~30$^{\circ}$ (to minimize dust extinction), have optical colors BP$-$RP = 0.55\,-\,0.90 (F5-G9 type, \citealt{PecautMamajek2013})\footnote{\url{https://www.pas.rochester.edu/~emamajek/EEM_dwarf_UBVIJHK_colors_Teff.txt}}, and have either large transverse velocities $v_t$~$>$~200~km\,s$^{-1}$ and/or large radial velocities $|v_r|$~$>$~200~km\,s$^{-1}$. We cross-match with the GALEX point-source catalog \citep{Bianchi2017,Bianchi2011} via the CDS XMatch service\footnote{\url{http://cdsxmatch.u-strasbg.fr/}}. We identify 15,088 (82$\%$) GALEX NUV counterparts within a tolerance of 5''. 

We compare the colors NUV$-$G versus BP$-$RP  of these 15,088 halo stars in Fig.~\ref{fig:HaloColorColor}. We overlay MIST models (red lines) of 10 Gyr old dwarf stars at [Fe/H] = $-$4, $-$2, and 0 \citep{Choi2016}. We display a reddening vector assuming a R$_{\rm V}$ = 3.1 dust-reddening law and the dust-extinction relations for different passbands from \citet{Schlafly2011}. 

We compute dust-corrected colors (BP$-$RP)$_{\rm o}$ and (NUV$-$G)$_{\rm o}$ based on the measured $E(g-r)$ from the 3D dust maps \citep{Green2019} and dust-extinction relations for different passbands assuming a Milky Way $R_V$ = 3.1 dust-reddening law (Table~6 in \citealt{Schlafly2011}). 
The optical color (BP$-$RP)$_{\rm o}$ maps the effective temperature while (NUV$-$G)$_{\rm o}$ traces metallicity. We therefore interpolate the theoretical MIST color (NUV$-$G)$_{\rm [Fe/H]=-4}$ of an [Fe/H]~=~$-$4 FG-dwarf as a function of (BP$-$RP)$_{\rm o}$.  We compute the difference:
\begin{equation}
\begin{split}
\label{eq:Dnuv}
   \Delta({\rm NUV}-{\rm G})_{\rm o;\small{[Fe/H]=-4}} = \\ ({\rm NUV}-{\rm G})_o-({\rm NUV}-{\rm G})_{\rm \small{[Fe/H]=-4}} 
\end{split}
\end{equation}

evaluated at the dust-corrected (BP$-$RP)$_{\rm o}$ color. A star with $\Delta$(NUV$-$G)$_{\rm o;[Fe/H]=-4}$ = 0 has the same (NUV$-$G)$_{\rm o}$ color as a theoretical MIST dwarf with [Fe/H] = $-$4.

Objects with smaller $\Delta$(NUV$-$G)$_{\rm o;[Fe/H]=-4}$ are closer to the [Fe/H] = $-$4 model and are therefore theoretically more likely to be a UMP star. The majority of halo stars are fainter in the NUV compared to the [Fe/H] = $-$2 MIST models. We select the 1,994 metal-poor candidates that exhibit an NUV excess compared to the [Fe/H] = $-$2 model and/or are within $\Delta$(NUV$-$G)$_{\rm [Fe/H]=-4}$ $\le$ 0.3 mag of the [Fe/H] = $-$4 model for the given BP$-$RP color (blue objects in Fig.~\ref{fig:HaloColorColor}).

We remove several types of artifacts in which the apparent UV excess is not due to a metal-poor stellar atmosphere. Using the {\it Gaia} archive, we search for bright companion stars within the GALEX NUV point spread function (5''), and we remove blended twin binaries and bright foreground/background stars. We remove systems with large astrometric errors, specifically systems with a Renormalized Unit Weight Error (RUWE) $>$ 2.5. RUWE represents the reduced chi-squared of {\it{Gaia}}'s 5-parameter astrometric solution, where a high RUWE value indicates the presence of an astrometric binary companion (\negthickspace\citealt{Gaia2023}). By removing stars with RUWE $>$ 2.5, we eliminate objects which likely contain a hidden astrometric binary companion that may bias the UV photometry. We cross-reference the metal-poor candidates with the Simbad database\footnote{\url{https://simbad.u-strasbg.fr/simbad/sim-fcoo}}, and we remove 14 known variable stars, as a periodic increase in flux during a GALEX observation would cause the star to falsely appear UV bright. We remove eight known spectroscopic binaries. We also cross-reference our sample with the Bayestar19 3D dust map \citep{Green2019} and remove objects with dust reddenings that exceed $E(g-r)$~$>$~0.20~mag.

\begin{deluxetable*}{rcccccccccc}[t!]
\label{tab:GaiaGalexProp}
\setlength{\tabcolsep}{3pt}
\tabletypesize{\footnotesize}
\tablehead{
\colhead{$\it{Gaia}$ ID} & 
\colhead{G} & 
\colhead{BP$-$RP} & 
\colhead{NUV} & 
\colhead{NUV$_{\rm err}$} & 
\colhead{$E(g-r)$} & 
\colhead{(BP$-$RP)$_{\rm o}$} & 
\colhead{(NUV$-$G)$_{\rm o}$} & 
\colhead{$\Delta$(NUV$-$G)$_{\rm o;[Fe/H]=-4}$} & 
\colhead{[Fe/H]} & 
\colhead{Reference}
}
\startdata
2314516094372555648 & 12.397 & 0.649 & 15.301 & 0.014 & 0.00 & 0.649 & 2.904 & 0.266 & -2.63 & Ryan \& Norris et. al. 1991 \\
\hline
2414078491471382912 & 13.358 & 0.602 & 16.114 & 0.019 & 0.00 & 0.602 & 2.756 & 0.272 & -2.44 & Limberg et. al. 2021 \\
\hline
2415821041306701312 & 12.781 & 0.631 & 15.516 & 0.011 & 0.00 & 0.631 & 2.735 & 0.157 & -2.53 & Limberg et. al. 2021 \\
\hline
2414995552888464256 & 13.960 & 0.633 & 16.774 & 0.019 & 0.01 & 0.620 & 2.764 & 0.220 & -2.29 & Limberg et. al. 2021 \\
\hline
2309460990880340864 & 12.665 & 0.643 & 15.408 & 0.008 & 0.00 & 0.643 & 2.743 & 0.126 & -4.09 & Soubiran et. al. 2022 \\
\hline
4905689556475416320 & 13.325 & 0.622 & 16.176 & 0.020 & 0.00 & 0.622 & 2.851 & 0.302 & -2.21 & Buder et. al. 2021 \\
\hline
2417057686946631296 & 13.408 & 0.773 & 16.850 & 0.019 & 0.00 & 0.773 & 3.442 & 0.266 & -1.83 & Limberg et. al. 2021 \\
\hline
2316326341484461440 & 13.641 & 0.613 & 16.466 & 0.012 & 0.02 & 0.587 & 2.722 & 0.287 & -1.82 & Limberg et. al. 2021 \\
\hline
2365100780371602560 & 14.443 & 0.581 & 17.139 & 0.030 & 0.00 & 0.581 & 2.696 & 0.280 & -2.86 & Limberg et. al. 2021 \\
\hline
2801174135693760256 & 14.358 & 0.677 & 17.247 & 0.023 & 0.07 & 0.587 & 2.536 & 0.099 & -2.82 & Limberg et. al. 2021 \\
\enddata
\caption{The first ten sources selected by the criteria described in Section \ref{sec:Sample}. A full table containing all 492 entries is available electronically. In the electronic version of the table, we also include each objects right ascension, declination, parallax, proper motion and transverse velocity. We list objects from our KOSMOS sample by marking their reference as ``KOSMOS".  The full list of references are below, in alphabetical order:\cite{Abdurrouf2022}, \cite{Aguado2019}, \cite{Amarsi2019}, \cite{Andales2024}, \cite{Barklem2005}, \cite{Beers1992}, \cite{Beers2017}, \cite{Boesgaard2011}, \cite{Bonifacio2019}, \cite{Bonifacio2024}, \cite{Buder2021}, \cite{Caffau2020}, \cite{Ceccarelli2024}, \cite{Chen2020}, \cite{Dietz2020}, \cite{Hourihane2023}, \cite{Ishigaki2013}, \cite{Jonsson2020}, \cite{Lai2008}, \cite{Li2022}, \cite{Limberg2021}, \cite{Lucchesi2022}, \cite{Mardini2024}, \cite{Placco2010}, \cite{Placco2022}, \cite{Roederer2014}, \cite{Ryan1991}, \cite{Roederer2014}, \cite{Sestito2020}, \cite{Shank2022}, \cite{Soubiran2016}, \cite{Soubiran2022}, \cite{Wang2021}, \cite{Yong2013}, \cite{Yong2021}, \cite{Zhang2023}, \cite{Zong2020}}
\end{deluxetable*}

We next search both the Simbad and VizieR\footnote{\url{https://vizier.unistra.fr/}} databases for literature references of spectroscopic metallicities, and we find 556 halo stars with previously measured [Fe/H]. We remove 74 objects with high metallicities [Fe/H] $>$ $-$1.5. These systems are mostly metal-rich stars that exhibit a UV excess due to being extremely chromospherically active. See Section~\ref{sec:Discussion} for further discussion of chromospheric activity and variability. Nonetheless, the population of metal-rich UV active stars is rather small. The contamination rate of metal-rich stars in our sample of halo stars that exhibit a NUV excess is only 74/556 = 13\%. 

We then perform our own observations and measure spectroscopic metallicities of 13 halo stars in our sample, of which 11 did not have previous metallicity measurements (see Sections~\ref{sec:Observations}\,-\,\ref{sec:abundances}). Our small addition of 11 stars brings the sample size to 493 halo stars with a NUV excess and measured [Fe/H] $<$ $-$1.5. We remove the single object with substantially negative $\Delta$(NUV$-$G)$_{\rm o;[Fe/H]=-4}$ $<$ $-$0.1~mag, which likely harbors a hot companion considering it is well below the [Fe/H] = $-$4 relation in Fig.~\ref{fig:HaloColorColor}. 

Stars brighter than NUV~$<$~15~mag enter the non-linear regime of the GALEX detector, and accurate photometry for sources brighter than NUV~$<$~13~mag are difficult to recover \citep{Morrissey2007,Camarota2014}. Fortunately, only 19 of our targets are brighter than NUV $<$ 15~mag, and none are brighter than NUV $<$ 13~mag. The vast majority (96\%) of our sample are in the linear regime and thus have accurate NUV photometry.

Our final statistical sample contains 492 halo stars with measured [Fe/H] $<$ $-$1.5 and $\Delta$(NUV$-$G)$_{\rm o;[Fe/H]=-4}$ $>$ $-$0.1 mag. In Table~\ref{tab:GaiaGalexProp}, we list their {\it Gaia} IDs, observed magnitudes and colors G, BP$-$RP, and NUV, dominant source of photometric error NUV$_{\rm err}$, dust-reddening $E(g-r)$, dust-corrected colors (BP$-$RP)$_{\rm o}$ and (NUV$-$G)$_{\rm o}$, photometric metallicity tracer $\Delta$(NUV$-$G)$_{\rm o;[Fe/H]=-4}$, spectroscopic metallicity [Fe/H], and corresponding reference. We discuss the statistical trends and properties of this sample in Section~\ref{sec:Discussion}.

\begin{deluxetable*}{rrrcccccccccccccccc}[t!]
\rotate
    \setlength{\tabcolsep}{3pt}
    \tablecaption{{\it Gaia}, GALEX, and 2MASS Properties of Observational KOSMOS Sample\label{tab:sample}}
    \tabletypesize{\footnotesize}
    \tablehead{
        \colhead{{\it Gaia} DR3 ID} & \colhead{RA~($^{\circ}$)} & \colhead{Dec~($^{\circ}$)} &  \colhead{$d$\,(pc)} & 
        \colhead{$v_{\rm t}$\,(km\,s$^{-1}$)} & \colhead{RUWE} &
        \colhead{G} & \colhead{BP$-$RP} & \colhead{J$-$K} & \colhead{NUV} & \colhead{$E(g-r)$} &
        \colhead{M$_{\rm G}$} & \colhead{(BP$-$RP)$_{\rm o}$} & 
        \colhead{(J$-$K)$_{\rm o}$} & 
        \colhead{(NUV$-$G)$_{\rm o}$} & 
        \colhead{$\Delta$(NUV$-$G)$_{\rm o;[Fe/H]=-4}$} & 
    }
    \startdata
    \hline
    2549908103516728704 & 11.790203 & 2.220088 & 421 & 236 & 1.07 & 12.61 & 0.58 & 0.29 & 15.23 & 0.04 & 4.40 & 0.53 & 0.27 & 2.42 & 0.12 \\ 
    \hline
    312013882108035840 & 12.690289 & 29.946421 & 507 & 278 & 1.59 & 13.22 & 0.75 & 0.36 & 16.36 & 0.09 & 4.50 & 0.64 & 0.32 & 2.69 & 0.08 \\ \hline
    20149516946735232 & 39.119036 & 8.734372 & 464 & 283 & 2.15 & 13.88 & 0.88 & 0.38 & 17.45 & 0.18 & 5.16 & 0.64 & 0.31 & 2.66 & 0.03 \\ \hline
    5129037304794991744 & 42.156189 & -19.441863 & 860 & 351 & 1.01 & 13.31 & 0.64 & 0.29 & 16.09 & 0.01 & 3.61 & 0.63 & 0.29 & 2.73 & 0.17 \\ \hline
    31296155375633152 & 47.653214 & 15.833620 & 394 & 256 & 0.97 & 12.87 & 0.89 & 0.39 & 16.81 & 0.15 & 4.56 & 0.69 & 0.33 & 3.19 & 0.35 \\ \hline
    60860511218929024 & 50.939342 & 20.142116 & 773 & 288 & 1.34 & 13.52 & 0.82 & 0.42 & 17.07 & 0.14 & 3.77 & 0.64 & 0.36 & 2.84 & 0.24 \\ \hline
    3187574373172628864 & 72.685106 & -7.554090 & 725 & 217 & 1.01 & 13.50 & 0.65 & 0.33 & 16.29 & 0.06 & 4.07 & 0.57 & 0.30 & 2.49 & 0.10 \\ \hline
    594226081365080960 & 139.784326 & 13.126845 & 387 & 382 & 1.01 & 12.44 & 0.65 & 0.34 & 15.30 & 0.05 & 4.39 & 0.59 & 0.32 & 2.61 & 0.20 \\ \hline
    1012126502345829120 & 139.814189 & 48.972107 & 628 & 342 & 0.95 & 12.55 & 0.66 & 0.30 & 15.36 & 0.05 & 3.45 & 0.60 & 0.28 & 2.56 & 0.09 \\ \hline
    793684774909242368 & 147.914124 & 33.629912 & 775 & 215 & 0.95 & 13.23 & 0.59 & 0.31 & 15.97 & 0.00 & 3.79 & 0.59 & 0.31 & 2.74 & 0.28 \\ \hline
    3967207625699469184 & 171.274913 & 15.211837 & 1020 & 465 & 0.92 & 14.31 & 0.62 & 0.30 & 17.02 & 0.00 & 4.26 & 0.62 & 0.30 & 2.71 & 0.19 \\ \hline
    2435684753950184448 & 354.988043 & -9.617601 & 361 & 322 & 1.45 & 11.95 & 0.65 & 0.35 & 14.89 & 0.01 & 4.17 & 0.63 & 0.29 & 2.89 & 0.30\\ \hline
    2739719922558093440 & 357.745346 & 2.603611 & 403 & 325 & 1.30 & 12.10 & 0.63 & 0.26 & 14.93 & 0.01 & 4.05 & 0.62 & 0.25 & 2.78 & 0.24 \\ \hline
    \enddata
\end{deluxetable*}

\section{Observations \& Data Reduction}
\label{sec:Observations}

We obtained long-slit spectra of 13 stars that satisfy our selection criteria. Only two had previously measured spectroscopic metallicities ($Gaia$ DR3 ID 594226081365080960 and 2739719922558093440). We list the {\it Gaia} and GALEX properties, magnitudes, and colors of our 13 selected stars in Table~\ref{tab:sample}. We also cross-match our 13 objects with the Two Micron All Sky Survey (2MASS; \citealt{Skrutskie2006}), and we list their J$-$K colors in Table~\ref{tab:sample}. Using the same procedures as for our full statistical sample, we compute dust-corrected colors and the metallicity tracer $\Delta$(NUV$-$G)$_{\rm o;[Fe/H]=-4}$. All 13 of our selected stars are late-F dwarfs with (BP$-$RP)$_{\rm o}$ = 0.53\,-\,0.69 and M$_{\rm G}$ = 3.5\,-\,5.2. Five of our targets have $\Delta$(NUV$-$G)$_{\rm o;[Fe/H]=-4}$ $<$
0.12~mag and therefore are potentially candidate EMP stars. 

We utilized the Kitt Peak Ohio State Multi-Object Spectrograph (KOSMOS) on the Astrophysical Research Consortium's (ARC) 3.5m telescope at Apache Point Observatory (APO)\footnote{\url{https://www.apo.nmsu.edu/arc35m/Instruments/KOSMOS/userguide.html}}. KOSMOS is a long-slit grism spectrograph with resolution R = 2,600. All observations were made with the blue grism and 0.87" slit at the center position, which covers 3,800\,-\,6,600\,\AA, thus spanning Ca K through H$\alpha$. We list our log of observations in Table~\ref{tab:obs}. We obtained calibration biases and flats during each night of observation. We acquired Argon lamp spectra for wavelength calibrations immediately after each science target.

\begin{deluxetable*}{rcccccc}[t!]
    \setlength{\tabcolsep}{6pt}
    \tablecaption{Observation Log of KOSMOS Targets\label{tab:obs}}
    \tabletypesize{\footnotesize}
    \tablehead{
        \colhead{$\it{Gaia}$ DR3 ID} & \colhead{UTC Date} & \colhead{UTC Time} & \colhead{Exp. Time} &
        \colhead{Seeing} & \colhead{Airmass} &
        \colhead{SNR (5500\AA)}
    }
    \startdata
    \hline
    2549908103516728704 & 2024-09-29 & 05:59 & 4\,$\times$\,360\,s & 1.4" & 1.2 & 123 \\
    \hline
    312013882108035840 & 2025-01-27 & 02:46 & 5\,$\times$\,300\,s & 1.7" & 1.3 & 106 \\
    \hline
    20149516946735232 & 2025-01-31 & 04:17 & 7\,$\times$\,300\,s & 1.2" & 1.6 & 104 \\
    \hline
    5129037304794991744 & 2025-01-27 & 03:36 & 6\,$\times$\,300\,s & 1.8" & 2.0 & 117 \\
    \hline
    31296155375633152 & 2025-02-02 & 04:26 & 6\,$\times$\,300\,s & 1.3" & 1.4 & 71 \\
    \hline
    60860511218929024 & 2025-03-01 & 03:58 & 6\,$\times$\,420\,s & 1.2" & 1.8 & 85 \\
    \hline
    3187574373172628864 & 2025-01-27 & 05:22 & 6\,$\times$\,300\,s & 1.8" & 1.5 & 105 \\
    \hline
    594226081365080960 & 2025-02-02 & 06:48 & 4\,$\times$\,300\,s & 1.4" & 1.1 & 119 \\
    \hline
    1012126502345829120 & 2025-01-27 & 06:23 & 4\,$\times$\,300\,s & 2.0" & 1.1 & 120 \\
    \hline
    793684774909242368 & 2025-01-31 & 06:01 & 5\,$\times$\,300\,s & 1.0" & 1.1 & 69 \\
    \hline
    3967207625699469184 & 2025-03-10 & 06:29 & 6\,$\times$\,420\,s & 1.1" & 1.1 & 98 \\
    \hline
    2435684753950184448 & 2025-07-26 & 08:51 & 3\,$\times$\,480\,s & 1.0" & 1.5 & 93 \\
    \hline
    2739719922558093440 & 2024-09-29 & 05:36 & 3\,$\times$\,360\,s & 1.4" & 1.2 & 153 \\
    \hline
    \enddata
\end{deluxetable*}

To reduce the raw spectra, we developed our own routine, the 
KOSMOS Astronomical Reduction Pipeline (KARP). KARP follows standard spectroscopic reduction procedures. It first median combines the calibration images to create a master bias and master flat. KARP then performs bias subtraction and flat fielding of each science exposure. KARP traces the central peak of the spectrum by fitting a Gaussian along each row. The centroid is then smoothed with a moving boxcar median to remove any artifacts. KARP extracts the spectrum by summing fluxes over a specified aperture centered on the trace and then subtracts an equivalent amount of sky background. The specified aperture width, buffer region, and sky background region can be adjusted to maximize the signal to noise ratio (SNR) for each object. For wavelength calibration, KARP fits vertical Gaussians to the profiles of bright Argon lines along the trace. KARP fits a fourth degree polynomial to the wavelength solution and then applies heliocentric velocity corrections.

\begin{figure}
    \centering
    \includegraphics[trim = 1.1cm 0.4cm 0.7cm 0.1cm, width=0.9\linewidth]{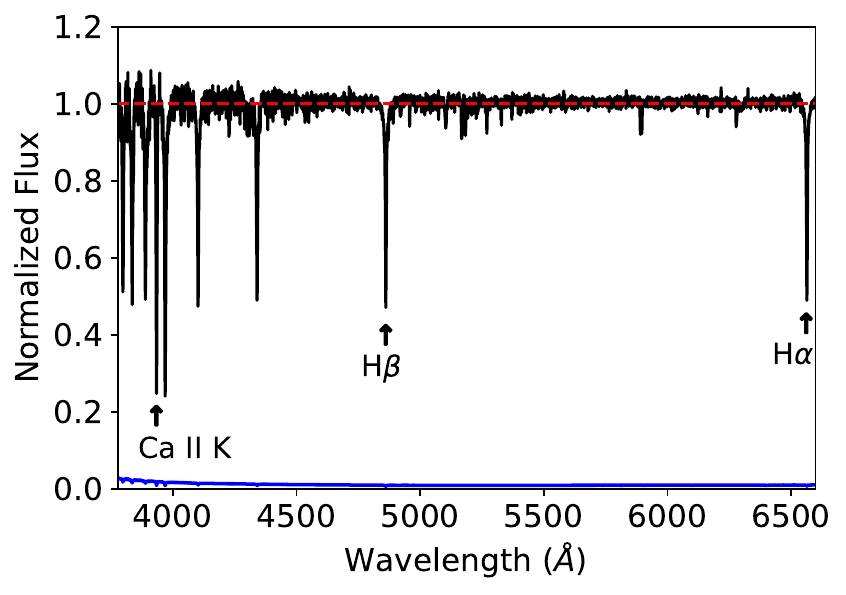}
    \caption{Normalized spectrum and error for {\it Gaia} ID 1012126502345829120. Note the deep Balmer lines (H$\alpha$ \& H$\beta$ are labeled) and Ca K. Several shallow iron features are visible across the spectrum.}
    \label{fig:NormSpec}
\end{figure}

KARP also normalizes each spectrum via a two-step process. First, it masks out deep absorption features (Balmer lines, Na D, Ca H\,\&\,K) and fits a fifth degree polynomial to the continuum. The normalized continuum after this first step still exhibits small undulations. Second, at each wavelength, KARP fits line segments over small wavelength regions, again masking over deep absorption features. The solution is then smoothed with a moving boxcar average, and the flux is then divided by this smoothed continuum. The normalized flux from each exposure is median combined into a final normalized spectrum. The aperture widths, sky background regions, smoothing lengths, etc. are varied until the SNR is maximized. We report the SNR at 5500\AA\ for each object in the final column of Table~\ref{tab:obs}.  We present an example normalized spectrum of one of our objects, {\it Gaia} ID 1012126502345829120, in Fig.~\ref{fig:NormSpec}. 

\begin{deluxetable*}{rrrrrrrrrrrrrrrr}[t!]
\rotate
    \label{tab:EW}
    \setlength{\tabcolsep}{4pt}
    \tablecaption{Equivalent Widths (\AA)}
    \tabletypesize{\footnotesize}
    \tablehead{
        \colhead{$\it{Gaia}$ DR3 ID} & \colhead{Ca II 3934} & \colhead{Fe I 4046} & \colhead{Fe I 4064} & \colhead{Ca I 4227} & \colhead{Fe I 4260} & \colhead{Fe I 4272} & \colhead{Fe I 4308} & \colhead{Fe I 4384} & \colhead{Fe I 4405} & \colhead{Fe I 4958} & \colhead{Mg I 5167} & \colhead{Mg I 5173} & \colhead{Mg I 5184} & \colhead{Fe I 5270} & \colhead{Fe I 5328}
    }
    \startdata
    \hline
    2549908103516728704 &        1.55 &        0.11 &        0.11 &        0.17 &        0.05 &        0.10 &        0.12 &        0.15 &        0.08 &        0.06 
                    &        0.13 &        0.15 &        0.16 &        0.10 &        0.09 \\
                    & $\pm$\,0.12 & $\pm$\,0.03 & $\pm$\,0.03 & $\pm$\,0.04 & $\pm$\,0.03 
                    & $\pm$\,0.03 & $\pm$\,0.03 & $\pm$\,0.04 & $\pm$\,0.02 & $\pm$\,0.02
                    & $\pm$\,0.02 & $\pm$\,0.03 & $\pm$\,0.02 & $\pm$\,0.02 & $\pm$\,0.02 \\
    \hline
    312013882108035840 &        2.07 &        0.16 &        0.10 &        0.22 &        0.07 &        0.11 &        0.16 &        0.19 &        0.12 &        0.13 
                   &        0.21 &        0.23 &        0.21 &        0.15 &        0.14 \\
                   & $\pm$\,0.11 & $\pm$\,0.04 & $\pm$\,0.04 & $\pm$\,0.04 & $\pm$\,0.04 
                   & $\pm$\,0.05 & $\pm$\,0.03 & $\pm$\,0.03 & $\pm$\,0.04 & $\pm$\,0.03
                   & $\pm$\,0.03 & $\pm$\,0.03 & $\pm$\,0.03 & $\pm$\,0.03 & $\pm$\,0.03 \\
    \hline
    20149516946735232 &        1.74 &        0.13 &        0.09 &        0.13 &        0.07 &        0.11 &        0.14 &        0.15 &        0.11 &        0.06 
                  &        0.16 &        0.14 &        0.18 &        0.08 &        0.07 \\
                  & $\pm$\,0.10 & $\pm$\,0.04 & $\pm$\,0.04 & $\pm$\,0.05 & $\pm$\,0.03 
                  & $\pm$\,0.03 & $\pm$\,0.03 & $\pm$\,0.03 & $\pm$\,0.03 & $\pm$\,0.02
                  & $\pm$\,0.03 & $\pm$\,0.03 & $\pm$\,0.03 & $\pm$\,0.02 & $\pm$\,0.02 \\
    \hline
    5129037304794991744 &        1.62 &        0.09 &        0.08 &        0.08 &        0.05 &        0.15 &        0.15 &        0.11 &        0.05 &        0.04 
                    &        0.16 &        0.15 &        0.16 &        0.09 &        0.10 \\
                    & $\pm$\,0.09 & $\pm$\,0.03 & $\pm$\,0.03 & $\pm$\,0.03 & $\pm$\,0.03 
                    & $\pm$\,0.03 & $\pm$\,0.03 & $\pm$\,0.03 & $\pm$\,0.03 & $\pm$\,0.02
                    & $\pm$\,0.03 & $\pm$\,0.03 & $\pm$\,0.03 & $\pm$\,0.02 & $\pm$\,0.02 \\
    \hline
    31296155375633152 &        5.02 &        0.30 &        0.18 &        0.44 &        0.13 &        0.21 &        0.23 &        0.23 &        0.13 &        0.20 
                  &        0.31 &        0.34 &        0.35 &        0.17 &        0.17 \\
                  & $\pm$\,0.10 & $\pm$\,0.09 & $\pm$\,0.05 & $\pm$\,0.07 & $\pm$\,0.04 
                  & $\pm$\,0.05 & $\pm$\,0.04 & $\pm$\,0.04 & $\pm$\,0.04 & $\pm$\,0.03
                  & $\pm$\,0.06 & $\pm$\,0.06 & $\pm$\,0.03 & $\pm$\,0.03 & $\pm$\,0.03 \\
    \hline
    60860511218929024 &        2.19 &        0.21 &        0.13 &        0.19 &        0.11 &        0.13 &        0.14 &        0.14 &        0.08 &        0.10 
                  &        0.19 &        0.22 &        0.19 &        0.09 &        0.09 \\
                  & $\pm$\,0.10 & $\pm$\,0.05 & $\pm$\,0.04 & $\pm$\,0.05 & $\pm$\,0.04 
                  & $\pm$\,0.04 & $\pm$\,0.04 & $\pm$\,0.04 & $\pm$\,0.03 & $\pm$\,0.03
                  & $\pm$\,0.05 & $\pm$\,0.04 & $\pm$\,0.03 & $\pm$\,0.03 & $\pm$\,0.03 \\
    \hline
    3187574373172628864 &        1.45 &        0.15 &        0.07 &        0.12 &        0.04 &        0.12 &        0.12 &        0.10 &        0.06 &        0.05 
                    &        0.12 &        0.16 &        0.15 &        0.11 &        0.11 \\
                    & $\pm$\,0.06 & $\pm$\,0.04 & $\pm$\,0.04 & $\pm$\,0.04 & $\pm$\,0.03 
                    & $\pm$\,0.03 & $\pm$\,0.03 & $\pm$\,0.03 & $\pm$\,0.03 & $\pm$\,0.03
                    & $\pm$\,0.04 & $\pm$\,0.04 & $\pm$\,0.03 & $\pm$\,0.03 & $\pm$\,0.03 \\
    \hline
    594226081365080960 &        2.29 &        0.09 &        0.12 &        0.18 &        0.05 &        0.14 &        0.14 &        0.15 &        0.10 &        0.10 
                   &        0.21 &        0.19 &        0.16 &        0.08 &        0.09 \\
                   & $\pm$\,0.15 & $\pm$\,0.04 & $\pm$\,0.04 & $\pm$\,0.04 & $\pm$\,0.04 
                   & $\pm$\,0.04 & $\pm$\,0.03 & $\pm$\,0.03 & $\pm$\,0.03 & $\pm$\,0.03
                   & $\pm$\,0.05 & $\pm$\,0.04 & $\pm$\,0.03 & $\pm$\,0.02 & $\pm$\,0.03 \\
    \hline
    1012126502345829120 &        3.11 &        0.17 &        0.15 &        0.24 &        0.07 &        0.15 &        0.16 &        0.18 &        0.12 &        0.13 
                   &        0.23 &        0.23 &        0.20 &        0.16 &        0.13 \\
                   & $\pm$\,0.15 & $\pm$\,0.04 & $\pm$\,0.04 & $\pm$\,0.05 & $\pm$\,0.03 
                   & $\pm$\,0.05 & $\pm$\,0.03 & $\pm$\,0.04 & $\pm$\,0.03 & $\pm$\,0.03
                   & $\pm$\,0.05 & $\pm$\,0.03 & $\pm$\,0.03 & $\pm$\,0.03 & $\pm$\,0.03 \\
    \hline
    793684774909242368 &        2.04 &        0.11 &        0.06 &        0.17 &        0.08 &        0.06 &        0.11 &        0.12 &        0.11 &        0.13 
                  &        0.19 &        0.14 &        0.14 &        0.17 &        0.10 \\
                  & $\pm$\,0.11 & $\pm$\,0.04 & $\pm$\,0.04 & $\pm$\,0.04 & $\pm$\,0.04 
                  & $\pm$\,0.04 & $\pm$\,0.05 & $\pm$\,0.04 & $\pm$\,0.04 & $\pm$\,0.03
                  & $\pm$\,0.04 & $\pm$\,0.04 & $\pm$\,0.04 & $\pm$\,0.03 & $\pm$\,0.03 \\
    \hline
    3967207625699469184 &        1.55 &        0.28 &        0.12 &        0.16 &        0.06 &        0.10 &        0.12 &        0.10 &        0.09 &        0.04 
                    &        0.15 &        0.13 &        0.15 &        0.11 &        0.07 \\
                    & $\pm$\,0.09 & $\pm$\,0.08 & $\pm$\,0.04 & $\pm$\,0.04 & $\pm$\,0.04 
                    & $\pm$\,0.04 & $\pm$\,0.04 & $\pm$\,0.03 & $\pm$\,0.03 & $\pm$\,0.03
                    & $\pm$\,0.04 & $\pm$\,0.04 & $\pm$\,0.03 & $\pm$\,0.03 & $\pm$\,0.03 \\
    \hline
    2435684753950184448 &        2.40 &        0.14 &        0.11 &        0.22 &        0.02 
                    &        0.24 &        0.15 &        0.20 &        0.09 &        0.12 
                    &        0.21 &        0.21 &        0.21 &        0.13 &        0.13 \\
                    & $\pm$\,0.06 & $\pm$\,0.03 & $\pm$\,0.06 & $\pm$\,0.04 & $\pm$\,0.05 
                    & $\pm$\,0.05 & $\pm$\,0.02 & $\pm$\,0.02 & $\pm$\,0.04 & $\pm$\,0.04
                    & $\pm$\,0.04 & $\pm$\,0.02 & $\pm$\,0.03 & $\pm$\,0.02 & $\pm$\,0.03 \\
    \hline
    2739719922558093440 &        1.34 &        0.09 &        0.07 &        0.12 &        0.03                       &       0.05 &        0.12 &        0.09 &        0.05 &        0.03 
                        &       0.11 &        0.09 &        0.11 &        0.05 &        0.05 \\
                        & $\pm$\,0.04 & $\pm$\,0.02 & $\pm$\,0.02 & $\pm$\,0.02 & $\pm$\,0.02 
                       & $\pm$\,0.01 & $\pm$\,0.02 & $\pm$\,0.02 & $\pm$\,0.01 & $\pm$\,0.01
                       & $\pm$\,0.02 & $\pm$\,0.02 & $\pm$\,0.02 & $\pm$\,0.02 & $\pm$\,0.01 \\
    \hline
    \hline
    \enddata
\end{deluxetable*}

\section{Abundances \& Physical Properties}
\label{sec:abundances}

To compute iron and alpha abundances, we first measure the equivalent widths (EWs) of 15 metal lines: 10 Fe lines spanning 4046\AA\ - 5328\AA, the Mg I triplet near 5173\AA, and both Ca II K (3934\AA) and Ca I 4227\AA. We list the wavelength centers of all 15 metal lines in the top row of Table~\ref{tab:EW}. We fit a Gaussian profile to each metal absorption line and integrate the Gaussian to compute the EW. We calculate both measurement uncertainties from the errors in the flux and systematic biases in the location of the normalized continuum. We add measurement and systematic errors in quadrature. We report the EWs and total errors of the 15 metal lines for all 13 stars in Table~\ref{tab:EW}.

We rely on deeper absorption lines to measure the radial velocities of our stars. Specifically, we fit Gaussian profiles to Ca II K (3934\AA) and the Doppler cores of six Balmer hydrogen lines: H$\alpha$, H$\gamma$, H$\delta$, H$\zeta$, H$\eta$, and H$\theta$. We exclude Ca II H (3968\AA) and H$\epsilon$ (3970\AA), which are blended with each other, and ignore H$\beta$, which lies in a region without Argon lines for reliable wavelength calibration and thus its velocity is more uncertain. We compute the weighted average and error of the radial velocities from the seven lines. We add the systematic error from our wavelength solution, typically 3\,km~s$^{-1}$, in quadrature with the measurement uncertainty. We report the average radial velocities and total errors for our 13 objects in Table~\ref{tab:PhysPropAbun}. 

\begin{deluxetable*}{rccrcccl}[t!]
    \setlength{\tabcolsep}{6pt}
    \tablecaption{Physical Properties and Abundances}
    \tabletypesize{\footnotesize}
    \label{tab:PhysPropAbun}
    \tablehead{
        \colhead{$\it{Gaia}$ DR3 ID} & \colhead{$T_{\rm eff}$\,(K)} & \colhead{log\,$g$} & \colhead{$v_{\rm r}$\,(km\,s$^{-1}$)} & \colhead{[Fe/H]} & \colhead{[$\alpha$/H]} & \colhead{[$\alpha$/Fe]} & \colhead{[Fe/H] Literature}
        }
    \startdata
    \hline
    2549908103516728704 & 6630 $\pm$ 110 & 4.5 & $-135 \pm 5$ & $-2.35 \pm 0.11$ & $-2.17 \pm 0.13$ & $0.18 \pm 0.16$ & - \\
    \hline
    312013882108035840 & 6190 $\pm$ 110 & 4.4 & $7 \pm 6$ & $-2.32 \pm 0.14$ & $-2.22 \pm 0.14$ & $0.09 \pm 0.14$ & - \\
    \hline
    20149516946735232 & 6200 $\pm$ 70 & 4.6 & $-144 \pm 8$ & $-2.64 \pm 0.12$ & $-2.49 \pm 0.10$ & $0.15 \pm 0.13$ & - \\
    \hline
    5129037304794991744 & 6280 $\pm$ 50 & 4.2 & $-16 \pm 9$ & $-2.59 \pm 0.12$ & $-2.42 \pm 0.14$ & $0.17 \pm 0.19$ & - \\
    \hline
    31296155375633152 & 6040 $\pm$ 80 & 4.4 & $-56 \pm 9$ & $-2.14 \pm 0.12$ & $-1.42 \pm 0.17$ & $0.73 \pm 0.12$ & - \\
    \hline
    60860511218929024 & 6160 $\pm$ 130 & 4.2 & $67 \pm 9$ & $-2.39 \pm 0.14$ & $-2.15 \pm 0.17$ & $0.24 \pm 0.16$ & - \\
    \hline
    3187574373172628864 & 6420 $\pm$ 130 & 4.2 & $312 \pm 8$ & $-2.50 \pm 0.16$ & $-2.34 \pm 0.16$ & $0.16 \pm 0.21$ & - \\
    \hline
    594226081365080960 & 6370 $\pm$ 140 & 4.3 & $78 \pm 8$ & $-2.43 \pm 0.14$ & $-2.09 \pm 0.17$ & $0.34 \pm 0.16$ & $-2.3$ \citep{Ryan1991} \\
    \hline
    1012126502345829120 & 6390 $\pm$ 70 & 4.1 & $59 \pm 6$ & $-1.97 \pm 0.13$ & $-1.49 \pm 0.15$ & $0.47 \pm 0.18$ & - \\
    \hline
    793684774909242368 & 6310 $\pm$ 110 & 4.2 & $127 \pm 10$ & $-2.31 \pm 0.20$ & $-2.19 \pm 0.11$ & $0.13 \pm 0.24$ & - \\
    \hline
    3967207625699469184 & 6300 $\pm$ 80 & 4.3 & $-99 \pm 8$ & $-2.52 \pm 0.15$ & $-2.47 \pm 0.08$ & $0.05 \pm 0.21$ & - \\
    \hline
    2435684753950184448 & 6330 $\pm$ 80 & 4.3 & $66 \pm 5$ & $-2.15 \pm 0.16$ & $-1.84 \pm 0.14$ & $0.31 \pm 0.18$ & - \\
    \hline
    2739719922558093440 & 6340 $\pm$ 40 & 4.2 & $33 \pm 5$ & $-2.92 \pm 0.10$ & $-2.48 \pm 0.06$ & $0.44 \pm 0.17$ & $-2.92$ \citep{Shen2023} \\
    \hline  
    \enddata
\end{deluxetable*}

We adopt surface gravities log\,$g$ based on {\it Gaia's} GSP-Phot Bayesian fits to photometry and XP spectra \citep{Andrae2023}. The GSP-Phot solutions for log\,$g$ are sufficiently reliable for our purposes, as the EWs of metal lines within our sample of solar-type dwarfs do not strongly depend on surface gravity, similar to other surveys of low-mass stars discussed in Section~\ref{sec:Intro}. All 13 stars are dwarfs with log\,$g$ = 4.1\,-\,4.6 (see Table~\ref{tab:PhysPropAbun}). 

Meanwhile, the EWs of metal lines strongly depend on effective temperature $T_{\rm eff}$. We therefore measure $T_{\rm eff}$ via four photometric methods. First, we rely on the MCMC samples of {\it Gaia's} GSP-Phot parameters, which generally show a degenerate solution between $T_{\rm eff}$ and $E$(BP$-$RP).  We estimate $T_{\rm eff}$ based on the measured $E$(BP$-$RP) determined from the 3D dust maps \citep{Green2019}. We next utilize the empirical relations in \citet{PecautMamajek2013}\footnote{\url{https://www.pas.rochester.edu/~emamajek/EEM_dwarf_UBVIJHK_colors_Teff.txt}} between both $T_{\rm eff}$ versus optical color (BP$-$RP)$_{\rm o}$ and $T_{\rm eff}$ versus near-infrared color (J$-$K)$_{\rm o}$. Finally, we interpolate $T_{\rm eff}$ versus (BP$-$RP)$_{\rm o}$ based on the MIST evolutionary tracks \citep{Choi2016}. We compute the average and standard deviation of the four methods, and we present the results in Table~\ref{tab:PhysPropAbun}. The temperatures of our late-F dwarfs span 6040\,-\,6630\,K with an average uncertainty of 90\,K. 

To measure abundances, we downloaded a grid of normalized AMBRE synthetic spectra \citep{deLaverny2012} from the POLLUX database \citep{Palacios2010}. The grid combinations include log\,$g$ = 4.0 and 4.5, $T_{\rm eff}$ = 6000\,K, 6250\,K, 6500\,K, and 6750\,K, and [Fe/H] = $-$4, $-$3, $-$2.5, $-$2, and $-$1.5, all scaled to solar abundances. We measure theoretical EWs of our 15 metal lines by numerically integrating the normalized AMBRE synthetic spectra. We create a denser grid of theoretical EWs with respect to effective temperature and metallicity. Specifically,  we bilinearly interpolate the theoretical EWs at intervals of 10\,K in temperature and 0.01 dex in abundance. The spacing of 0.5 dex in surface gravity is already sufficient that we do not need to interpolate onto a finer grid. We match the measured EWs to the theoretical EWs with the closest $T_{\rm eff}$ and log\,$g$. In this manner, the measured EW of each metal line gives a unique solution for [X$_i$/H]. We propagate the uncertainties in the measured EWs into uncertainties for [X$_i$/H]. We then compute a weighted average and error of [Fe/H] based on the 10 Fe lines and a weighted average and error of [$\alpha$/H] based on the three Mg I lines and the two Ca lines. Finally, we propagate the uncertainties in effective temperature by repeating the process above but evaluated at the measured $\pm$1$\sigma$ errors in $T_{\rm eff}$. We add the measurement errors due to the uncertainties in the EWs and the systematic error due to the uncertainty in $T_{\rm eff}$ in quadrature. In general, the uncertainties in [Fe/H] and [$\alpha$/H] are dominated by the error in $T_{\rm eff}$ while the computed ratio [$\alpha$/Fe] is rather insensitive to temperature. We report the weighted averages and total errors of [Fe/H], [$\alpha$/H], and [$\alpha$/Fe] for our 13 stars in Table~\ref{tab:PhysPropAbun}.

All but one of our 13 stars are VMP stars with $-$2.9 $<$ [Fe/H] $<$ $-$2.0. Unfortunately, none of our observed targets are EMP stars, even the few with $\Delta$(NUV$-$G)$_{\rm o;[Fe/H]=-4}$ $<$ 0.1 (see Section~\ref{sec:Discussion} for discussion of the full statistical sample). All 13 of our stars are $\alpha$ enhanced with [$\alpha$/Fe] = 0.1\,-\,0.7. For the two objects with previously measured metallicities, our value of [Fe/H] = $-$2.43\,$\pm$\,0.14 for {\it Gaia} ID 594226081365080960 is consistent with the literature value of [Fe/H] = $-$2.3\,$\pm$\,0.2 \citep{Ryan1991}, and our measurement of [Fe/H] = $-$2.92\,$\pm$\,0.10 for {\it Gaia} ID 2739719922558093440 exactly matches the previous result of [Fe/H] = $-$2.92\,$\pm$\,0.02 \citep{Shen2023}. This consistency helps to verify our method of measuring abundances from EWs and photometric temperatures.

\section{Discussion}
\label{sec:Discussion}
We add our observed KOSMOS targets to the literature sample with previously measured spectroscopic metallicities, bringing the total sample size to 492 objects (Table~\ref{tab:GaiaGalexProp}). In our full sample, 54 have $-$2~$<$~[Fe/H]~$<$~$-$1.5, 407 are VMP stars with $-$3~$<$~[Fe/H]~$\le$~$-$2, 29 are EMP stars with $-$4~$<$~[Fe/H]~$\le$~$-$3, and only 2 are UMP stars with [Fe/H]~$\le$~$-$4. According to the {\it Gaia} GSP-Phot parameters, the vast majority (98\%) of our sample are dwarfs with log\,$g$~=~3.8\,-\,4.6.  The metallicity tracer $\Delta$(NUV$-$G)$_{\rm o;[Fe/H]=-4}$ does not depend significantly on surface gravity. For example, a 5,800\,K PHOENIX model spectrum with log\,$g$ = 3.5 exhibits the same NUV excess (within 0.01 mag) as the log\,$g$ = 4.5 model in Fig.~\ref{fig:Phoenix}. We also do not find any significant differences in the distributions of $\Delta$(NUV$-$G)$_{\rm o;[Fe/H]=-4}$ between our dwarfs with log\,$g$ = 3.8\,-\,4.2 versus log\,$g$ = 4.2\,-\,4.6.

\begin{figure}
    \centering
    \includegraphics[trim = 1.5cm 0.8cm 0.8cm 0cm, width=0.9\linewidth]{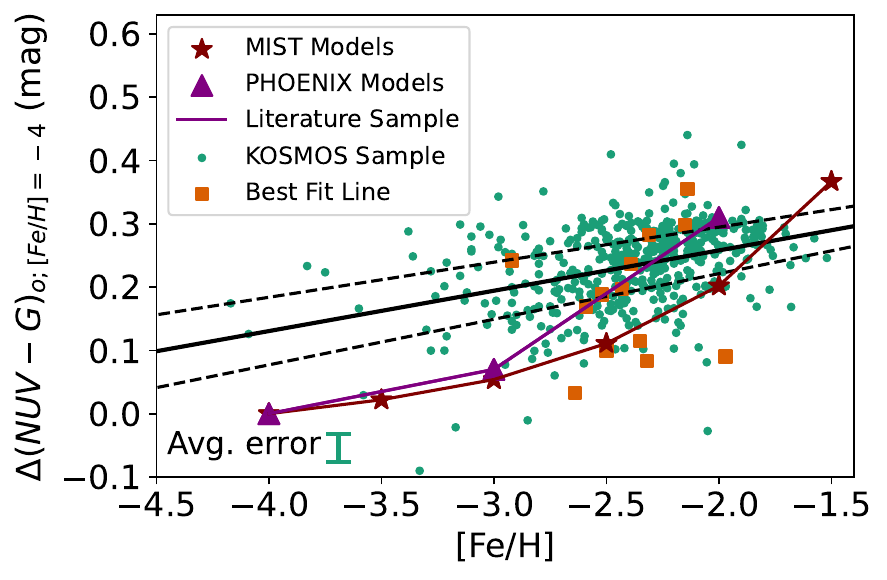}
    \caption{Photometric metallicity tracer $\Delta$(NUV$-$G)$_{\rm o;[Fe/H]=-4}$ versus actual spectroscopic metallicities [Fe/H] for the 492 stars in our full sample (green), including the 13 in our KOSMOS sample (orange). Our best-fit line (solid) and error (dashed) are shown in black, theoretical MIST models are shown in red, and theoretical PHOENIX models are shown in purple. Average GALEX NUV error is shown with a light green bar.}
    \label{fig:Delm4_FeH_all}
\end{figure}

We compare the photometric metallicity tracer $\Delta$(NUV$-$G)$_{\rm o;[Fe/H]=-4}$ to their actual spectroscopic metallicities [Fe/H] in Fig.~\ref{fig:Delm4_FeH_all}. We also overlay the theoretical relation from MIST models for the average optical color (BP$-$RP)$_{\rm o}$ = 0.67 of our sample. By definition, the theoretical metallicity tracer is $\Delta$(NUV$-$G)$_{\rm o;[Fe/H]=-4}$ = 0.0 at [Fe/H] = $-$4. 

We indeed find a visibly positive correlation between $\Delta$(NUV$-$G)$_{\rm o;[Fe/H]=-4}$ and [Fe/H]. We fit a linear relation (black line in Fig.~\ref{fig:Delm4_FeH_all}) and measure a positive slope that is discrepant with zero at the 8$\sigma$ level. We measure a Spearman rank correlation coefficient of 0.39 between $\Delta$(NUV$-$G)$_{o;[Fe/H]=-4}$ and [Fe/H] with a probability of no correlation of $p$ = 10$^{-18}$ (8$\sigma$). The photometric metallicity tracer $\Delta$(NUV$-$G)$_{\rm o;[Fe/H]=-4}$ and actual spectroscopic metallicities [Fe/H] are positively correlated at a statistically significant level. 

However, the data are inconsistent with the MIST models. On average, VMP stars with $-$3~$<$~[Fe/H]~$<$~$-$2 and especially EMP stars with $-$4~$<$~[Fe/H]~$<$~$-$3 are fainter in the NUV compared to the theoretical MIST models. The MIST colors are based on \citet{Kurucz1993} ATLAS12/SYNTHE synthetic atmospheres \citep{Choi2016}, which appear 0.1~mag brighter in the NUV compared to our actual metal-poor halo stars. Meanwhile, the PHOENIX models more reliably match the bulk trend of our sample, especially near [Fe/H] = $-$2.5.

Moreover, the intrinsic scatter in $\Delta$(NUV$-$G)$_{\rm o;[Fe/H]=-4}$ greatly exceeds the measurement uncertainties. The rms dispersion around our best-fit line is 0.062 mag, substantially higher than the average error of 0.022 mag in the NUV. Systematic errors in dust-reddening corrections or contamination by blended binaries likely play only a minor role in increasing the scatter of $\Delta$(NUV$-$G)$_{\rm o;[Fe/H]=-4}$. We instead conclude that the NUV fluxes of metal-poor halo stars likely exhibit intrinsic scatter due to chromospheric activity, both temporal variability and variations from object to object. Solar-type stars can vary up to 20\% in the UV during a single rotation period, while long-term magnetic cycles can change the UV flux by an additional 5\%\,-\,10\% \citep{Sofia1989,Deland2012,Sreejith2020}. These UV variations are consistent with the observed scatter in our $\Delta$(NUV$-$G)$_{\rm o;[Fe/H]=-4}$ diagnostic.

The large scatter in NUV inhibits our ability to distinguish VMP stars from EMP stars. A halo star with $\Delta$(NUV$-$G)$_{\rm o;[Fe/H]=-4}$ = 0.1 is more likely to be a UV active VMP star than a quiescent EMP star. Nonetheless, our 
metallicity tracer $\Delta$(NUV$-$G)$_{\rm o;[Fe/H]=-4}$ can reliably separate VMP stars from their metal-rich counterparts. For example, the fraction of stars with [Fe/H]~$<$~$-$2 compared to those with [Fe/H]~$<$~$-$1.5 increases from 88\% for $\Delta$(NUV$-$G)$_{\rm o;[Fe/H]=-4}$ $<$ 0.3 to 98\% for $\Delta$(NUV$-$G)$_{\rm o;[Fe/H]=-4}$ $<$ 0.15. The metallicity tracer yields a relatively pure sample of VMP and EMP stars. 

\section{Conclusions}
\label{sec:conclusions}
We summarize our main results as follows:

\begin{enumerate}
    \item We compiled a sample of 492 bright halo stars with GALEX NUV photometry, {\it Gaia} data, and ground-based spectroscopic metallicities.  We utilized the KOSMOS spectrograph at APO to measure iron and $\alpha$ abundances for 13 of these stars. We find a statistically significant correlation between our photometric metallicity tracer $\Delta(NUV-G)_{\rm o;[Fe/H]=-4}$ and spectroscopic [Fe/H]. We measure a Spearman rank correlation coefficient of 0.39 between $\Delta$(NUV$-$G)$_{\rm o;[Fe/H]=-4}$ and [Fe/H], with a probability of no correlation of $p$ = 10$^{-18}$ (8$\sigma$).
    \item The intrinsic scatter in $\Delta$(NUV$-$G)$_{\rm o;[Fe/H]=-4}$ greatly exceeds the measurement uncertainties. We conclude that the NUV fluxes of metal-poor halo stars exhibit intrinsic scatter due to chromospheric activity, from both temporal variability and variations from object to object. Nearby stars also exhibit NUV variability of 10\%\,-\,20\% due to rotational effects of starspots and long-term magnetic cycle activity.
    \item All of the targeted stars in our KOSMOS sample have [Fe/H]~$<$~$-$1.97, confirming that VMP stars can be reliably identified based on their GALEX NUV excess. While it is possible to differentiate VMP stars from their metal-rich counterparts based on GALEX NUV photometry, intrinsic NUV chromospheric variability does not allow for the separation of EMP stars from VMP stars.
\end{enumerate}

This work was funded by NASA ADAP grant 80NSSC25K7586, NASA EPSCoR grant 80NSSC22M0053, and NSF REU grant AST 1852289. We thank the APO telescope and technical staff for their assistance during our observations.

\clearpage

\bibliographystyle{aasjournalv7}                       
\bibliography{biblio}

@ARTICLE{Morrissey2007,
       author = {{Morrissey}, Patrick and {Conrow}, Tim and {Barlow}, Tom A. and {Small}, Todd and {Seibert}, Mark and {Wyder}, Ted K. and {Budav{\'a}ri}, Tam{\'a}s and {Arnouts}, Stephane and {Friedman}, Peter G. and {Forster}, Karl and {Martin}, D. Christopher and {Neff}, Susan G. and {Schiminovich}, David and {Bianchi}, Luciana and {Donas}, Jos{\'e} and {Heckman}, Timothy M. and {Lee}, Young-Wook and {Madore}, Barry F. and {Milliard}, Bruno and {Rich}, R. Michael and {Szalay}, Alex S. and {Welsh}, Barry Y. and {Yi}, Sukyoung K.},
        title = "{The Calibration and Data Products of GALEX}",
      journal = {\apjs},
         year = 2007,
        month = dec,
       volume = {173},
       number = {2},
        pages = {682-697},
          doi = {10.1086/520512},
archivePrefix = {arXiv},
       eprint = {0706.0755},
 primaryClass = {astro-ph},
       adsurl = {https://ui.adsabs.harvard.edu/abs/2007ApJS..173..682M}
}

@ARTICLE{Abdurrouf2022,
       author = {{Abdurro'uf} and {Accetta}, Katherine and {Aerts}, Conny and {Silva Aguirre}, V{\'\i}ctor and {Ahumada}, Romina and {Ajgaonkar}, Nikhil and {Filiz Ak}, N. and {Alam}, Shadab and {Allende Prieto}, Carlos and {Almeida}, Andr{\'e}s and {Anders}, Friedrich and {Anderson}, Scott F. and {Andrews}, Brett H. and {Anguiano}, Borja and {Aquino-Ort{\'\i}z}, Erik and {Arag{\'o}n-Salamanca}, Alfonso and {Argudo-Fern{\'a}ndez}, Maria and {Ata}, Metin and {Aubert}, Marie and {Avila-Reese}, Vladimir and {Badenes}, Carles and {Barb{\'a}}, Rodolfo H. and {Barger}, Kat and {Barrera-Ballesteros}, Jorge K. and {Beaton}, Rachael L. and {Beers}, Timothy C. and {Belfiore}, Francesco and {Bender}, Chad F. and {Bernardi}, Mariangela and {Bershady}, Matthew A. and {Beutler}, Florian and {Bidin}, Christian Moni and {Bird}, Jonathan C. and {Bizyaev}, Dmitry and {Blanc}, Guillermo A. and {Blanton}, Michael R. and {Boardman}, Nicholas Fraser and {Bolton}, Adam S. and {Boquien}, M{\'e}d{\'e}ric and {Borissova}, Jura and {Bovy}, Jo and {Brandt}, W.~N. and {Brown}, Jordan and {Brownstein}, Joel R. and {Brusa}, Marcella and {Buchner}, Johannes and {Bundy}, Kevin and {Burchett}, Joseph N. and {Bureau}, Martin and {Burgasser}, Adam and {Cabang}, Tuesday K. and {Campbell}, Stephanie and {Cappellari}, Michele and {Carlberg}, Joleen K. and {Wanderley}, F{\'a}bio Carneiro and {Carrera}, Ricardo and {Cash}, Jennifer and {Chen}, Yan-Ping and {Chen}, Wei-Huai and {Cherinka}, Brian and {Chiappini}, Cristina and {Choi}, Peter Doohyun and {Chojnowski}, S. Drew and {Chung}, Haeun and {Clerc}, Nicolas and {Cohen}, Roger E. and {Comerford}, Julia M. and {Comparat}, Johan and {da Costa}, Luiz and {Covey}, Kevin and {Crane}, Jeffrey D. and {Cruz-Gonzalez}, Irene and {Culhane}, Connor and {Cunha}, Katia and {Dai}, Y. Sophia and {Damke}, Guillermo and {Darling}, Jeremy and {Davidson}, Jr., James W. and {Davies}, Roger and {Dawson}, Kyle and {De Lee}, Nathan and {Diamond-Stanic}, Aleksandar M. and {Cano-D{\'\i}az}, Mariana and {S{\'a}nchez}, Helena Dom{\'\i}nguez and {Donor}, John and {Duckworth}, Chris and {Dwelly}, Tom and {Eisenstein}, Daniel J. and {Elsworth}, Yvonne P. and {Emsellem}, Eric and {Eracleous}, Mike and {Escoffier}, Stephanie and {Fan}, Xiaohui and {Farr}, Emily and {Feng}, Shuai and {Fern{\'a}ndez-Trincado}, Jos{\'e} G. and {Feuillet}, Diane and {Filipp}, Andreas and {Fillingham}, Sean P. and {Frinchaboy}, Peter M. and {Fromenteau}, Sebastien and {Galbany}, Llu{\'\i}s and {Garc{\'\i}a}, Rafael A. and {Garc{\'\i}a-Hern{\'a}ndez}, D.~A. and {Ge}, Junqiang and {Geisler}, Doug and {Gelfand}, Joseph and {G{\'e}ron}, Tobias and {Gibson}, Benjamin J. and {Goddy}, Julian and {Godoy-Rivera}, Diego and {Grabowski}, Kathleen and {Green}, Paul J. and {Greener}, Michael and {Grier}, Catherine J. and {Griffith}, Emily and {Guo}, Hong and {Guy}, Julien and {Hadjara}, Massinissa and {Harding}, Paul and {Hasselquist}, Sten and {Hayes}, Christian R. and {Hearty}, Fred and {Hern{\'a}ndez}, Jes{\'u}s and {Hill}, Lewis and {Hogg}, David W. and {Holtzman}, Jon A. and {Horta}, Danny and {Hsieh}, Bau-Ching and {Hsu}, Chin-Hao and {Hsu}, Yun-Hsin and {Huber}, Daniel and {Huertas-Company}, Marc and {Hutchinson}, Brian and {Hwang}, Ho Seong and {Ibarra-Medel}, H{\'e}ctor J. and {Chitham}, Jacob Ider and {Ilha}, Gabriele S. and {Imig}, Julie and {Jaekle}, Will and {Jayasinghe}, Tharindu and {Ji}, Xihan and {Johnson}, Jennifer A. and {Jones}, Amy and {J{\"o}nsson}, Henrik and {Katkov}, Ivan and {Khalatyan}, Dr., Arman and {Kinemuchi}, Karen and {Kisku}, Shobhit and {Knapen}, Johan H. and {Kneib}, Jean-Paul and {Kollmeier}, Juna A. and {Kong}, Miranda and {Kounkel}, Marina and {Kreckel}, Kathryn and {Krishnarao}, Dhanesh and {Lacerna}, Ivan and {Lane}, Richard R. and {Langgin}, Rachel and {Lavender}, Ramon and {Law}, David R. and {Lazarz}, Daniel and {Leung}, Henry W. and {Leung}, Ho-Hin and {Lewis}, Hannah M. and {Li}, Cheng and {Li}, Ran and {Lian}, Jianhui and {Liang}, Fu-Heng and {Lin}, Lihwai and {Lin}, Yen-Ting and {Lin}, Sicheng and {Lintott}, Chris and {Long}, Dan and {Longa-Pe{\~n}a}, Pen{\'e}lope and {L{\'o}pez-Cob{\'a}}, Carlos and {Lu}, Shengdong and {Lundgren}, Britt F. and {Luo}, Yuanze and {Mackereth}, J. Ted and {de la Macorra}, Axel and {Mahadevan}, Suvrath and {Majewski}, Steven R. and {Manchado}, Arturo and {Mandeville}, Travis and {Maraston}, Claudia and {Margalef-Bentabol}, Berta and {Masseron}, Thomas and {Masters}, Karen L. and {Mathur}, Savita and {McDermid}, Richard M. and {Mckay}, Myles and {Merloni}, Andrea and {Merrifield}, Michael and {Meszaros}, Szabolcs and {Miglio}, Andrea and {Di Mille}, Francesco and {Minniti}, Dante and {Minsley}, Rebecca and {Monachesi}, Antonela},
        title = "{The Seventeenth Data Release of the Sloan Digital Sky Surveys: Complete Release of MaNGA, MaStar, and APOGEE-2 Data}",
      journal = {\apjs},
         year = 2022,
        month = apr,
       volume = {259},
       number = {2},
          eid = {35},
        pages = {35},
          doi = {10.3847/1538-4365/ac4414},
archivePrefix = {arXiv},
       eprint = {2112.02026},
 primaryClass = {astro-ph.GA},
       adsurl = {https://ui.adsabs.harvard.edu/abs/2022ApJS..259...35A}
}

@ARTICLE{Aguado2019,
       author = {{Aguado}, David S. and {Youakim}, Kris and {Gonz{\'a}lez Hern{\'a}ndez}, Jonay I. and {Allende Prieto}, Carlos and {Starkenburg}, Else and {Martin}, Nicolas and {Bonifacio}, Piercarlo and {Arentsen}, Anke and {Caffau}, Elisabetta and {Peralta de Arriba}, Luis and {Sestito}, Federico and {Garcia-Dias}, Rafael and {Fantin}, Nicholas and {Hill}, Vanessa and {Jablonca}, Pascale and {Jahandar}, Farbod and {Kielty}, Collin and {Longeard}, Nicolas and {Lucchesi}, Romain and {S{\'a}nchez-Janssen}, Rub{\'e}n and {Osorio}, Yeisson and {Palicio}, Pedro A. and {Tolstoy}, Eline and {Wilson}, Thomas G. and {C{\^o}t{\'e}}, Patrick and {Kordopatis}, Georges and {Lardo}, Carmela and {Navarro}, Julio F. and {Thomas}, Guillaume F. and {Venn}, Kim},
        title = "{The Pristine survey - VI. The first three years of medium-resolution follow-up spectroscopy of Pristine EMP star candidates}",
      journal = {\mnras},
         year = 2019,
        month = dec,
       volume = {490},
       number = {2},
        pages = {2241-2253},
          doi = {10.1093/mnras/stz2643},
archivePrefix = {arXiv},
       eprint = {1909.08138},
 primaryClass = {astro-ph.GA},
       adsurl = {https://ui.adsabs.harvard.edu/abs/2019MNRAS.490.2241A}
}

@ARTICLE{Amarsi2019,
       author = {{Amarsi}, A.~M. and {Nissen}, P.~E. and {Sk{\'u}lad{\'o}ttir}, {\'A}.},
        title = "{Carbon, oxygen, and iron abundances in disk and halo stars. Implications of 3D non-LTE spectral line formation}",
      journal = {\aap},
         year = 2019,
        month = oct,
       volume = {630},
          eid = {A104},
        pages = {A104},
          doi = {10.1051/0004-6361/201936265},
archivePrefix = {arXiv},
       eprint = {1908.10319},
 primaryClass = {astro-ph.SR},
       adsurl = {https://ui.adsabs.harvard.edu/abs/2019A&A...630A.104A}
}

@ARTICLE{Andales2024,
       author = {{Andales}, Hillary Diane and {Santos Figueiredo}, Ananda and {Fienberg}, Casey Gordon and {Mardini}, Mohammad K. and {Frebel}, Anna},
        title = "{The oldest stars with low neutron-capture element abundances and origins in ancient dwarf galaxies}",
      journal = {\mnras},
         year = 2024,
        month = jun,
       volume = {530},
       number = {4},
        pages = {4712-4729},
          doi = {10.1093/mnras/stae670},
archivePrefix = {arXiv},
       eprint = {2405.07856},
 primaryClass = {astro-ph.GA},
       adsurl = {https://ui.adsabs.harvard.edu/abs/2024MNRAS.530.4712A}
}

@ARTICLE{Barklem2005,
       author = {{Barklem}, P.~S. and {Christlieb}, N. and {Beers}, T.~C. and {Hill}, V. and {Bessell}, M.~S. and {Holmberg}, J. and {Marsteller}, B. and {Rossi}, S. and {Zickgraf}, F. -J. and {Reimers}, D.},
        title = "{The Hamburg/ESO R-process enhanced star survey (HERES). II. Spectroscopic analysis of the survey sample}",
      journal = {\aap},
         year = 2005,
        month = aug,
       volume = {439},
       number = {1},
        pages = {129-151},
          doi = {10.1051/0004-6361:20052967},
archivePrefix = {arXiv},
       eprint = {astro-ph/0505050},
 primaryClass = {astro-ph},
       adsurl = {https://ui.adsabs.harvard.edu/abs/2005A&A...439..129B}
}

@ARTICLE{Beers2017,
       author = {{Beers}, Timothy C. and {Placco}, Vinicius M. and {Carollo}, Daniela and {Rossi}, Silvia and {Lee}, Young Sun and {Frebel}, Anna and {Norris}, John E. and {Dietz}, Sarah and {Masseron}, Thomas},
        title = "{Bright Metal-Poor Stars from the Hamburg/ESO Survey. II. A Chemodynamical Analysis}",
      journal = {\apj},
         year = 2017,
        month = jan,
       volume = {835},
       number = {1},
          eid = {81},
        pages = {81},
          doi = {10.3847/1538-4357/835/1/81},
archivePrefix = {arXiv},
       eprint = {1611.03762},
 primaryClass = {astro-ph.SR},
       adsurl = {https://ui.adsabs.harvard.edu/abs/2017ApJ...835...81B}
}

@ARTICLE{Beers1992,
       author = {{Beers}, Timothy C. and {Preston}, George W. and {Shectman}, Stephen A.},
        title = "{A Search for Stars of Very Low Metal Abundance. II}",
      journal = {\aj},
         year = 1992,
        month = jun,
       volume = {103},
        pages = {1987},
          doi = {10.1086/116207},
       adsurl = {https://ui.adsabs.harvard.edu/abs/1992AJ....103.1987B}
}

@ARTICLE{Boesgaard2011,
       author = {{Boesgaard}, Ann Merchant and {Rich}, Jeffrey A. and {Levesque}, Emily M. and {Bowler}, Brendan P.},
        title = "{Beryllium and Alpha-element Abundances in a Large Sample of Metal-poor Stars}",
      journal = {\apj},
         year = 2011,
        month = dec,
       volume = {743},
       number = {2},
          eid = {140},
        pages = {140},
          doi = {10.1088/0004-637X/743/2/140},
archivePrefix = {arXiv},
       eprint = {1110.2823},
 primaryClass = {astro-ph.SR},
       adsurl = {https://ui.adsabs.harvard.edu/abs/2011ApJ...743..140B}
}

@ARTICLE{Bianchi2017,
       author = {{Bianchi}, Luciana and {Shiao}, Bernie and {Thilker}, David},
        title = "{Revised Catalog of GALEX Ultraviolet Sources. I. The All-Sky Survey: GUVcat\_AIS}",
      journal = {\apjs},
         year = 2017,
        month = jun,
       volume = {230},
       number = {2},
          eid = {24},
        pages = {24},
          doi = {10.3847/1538-4365/aa7053},
archivePrefix = {arXiv},
       eprint = {1704.05903},
 primaryClass = {astro-ph.GA},
       adsurl = {https://ui.adsabs.harvard.edu/abs/2017ApJS..230...24B}
}

@ARTICLE{Bianchi2011,
       author = {{Bianchi}, L. and {Herald}, J. and {Efremova}, B. and {Girardi}, L. and {Zabot}, A. and {Marigo}, P. and {Conti}, A. and {Shiao}, B.},
        title = "{GALEX catalogs of UV sources: statistical properties and sample science applications: hot white dwarfs in the Milky Way}",
      journal = {\apss},
         year = 2011,
        month = sep,
       volume = {335},
       number = {1},
        pages = {161-169},
          doi = {10.1007/s10509-010-0581-x},
       adsurl = {https://ui.adsabs.harvard.edu/abs/2011Ap&SS.335..161B}
}

@ARTICLE{Bonifacio2019,
       author = {{Bonifacio}, P. and {Caffau}, E. and {Sestito}, F. and {Lardo}, C. and {Martin}, N.~F. and {Starkenburg}, E. and {Sbordone}, L. and {Fran{\c{c}}ois}, P. and {Jablonka}, P. and {Henden}, A.~A. and {Salvadori}, S. and {Gonz{\'a}lez Hern{\'a}ndez}, J.~I. and {Aguado}, D.~S. and {Hill}, V. and {Venn}, K. and {Navarro}, J.~F. and {Arentsen}, A. and {Sanchez-Janssen}, R. and {Carlberg}, R.},
        title = "{The Pristine survey - V. A bright star sample observed with SOPHIE}",
      journal = {\mnras},
         year = 2019,
        month = aug,
       volume = {487},
       number = {3},
        pages = {3797-3814},
          doi = {10.1093/mnras/stz1378},
       adsurl = {https://ui.adsabs.harvard.edu/abs/2019MNRAS.487.3797B}
}

@ARTICLE{Bonifacio2024,
       author = {{Bonifacio}, P. and {Caffau}, E. and {Monaco}, L. and {Sbordone}, L. and {Spite}, M. and {Mucciarelli}, A. and {Fran{\c{c}}ois}, P. and {Lombardo}, L. and {Matas Pinto}, A. d. M.},
        title = "{High-speed stars. II. An unbound star, young stars, bulge metal-poor stars, and Aurora candidates}",
      journal = {\aap},
         year = 2024,
        month = apr,
       volume = {684},
          eid = {A91},
        pages = {A91},
          doi = {10.1051/0004-6361/202347865},
archivePrefix = {arXiv},
       eprint = {2402.02876},
 primaryClass = {astro-ph.GA},
       adsurl = {https://ui.adsabs.harvard.edu/abs/2024A&A...684A..91B}
}

@ARTICLE{Buder2021,
       author = {{Buder}, Sven and {Sharma}, Sanjib and {Kos}, Janez and {Amarsi}, Anish M. and {Nordlander}, Thomas and {Lind}, Karin and {Martell}, Sarah L. and {Asplund}, Martin and {Bland-Hawthorn}, Joss and {Casey}, Andrew R. and {de Silva}, Gayandhi M. and {D'Orazi}, Valentina and {Freeman}, Ken C. and {Hayden}, Michael R. and {Lewis}, Geraint F. and {Lin}, Jane and {Schlesinger}, Katharine J. and {Simpson}, Jeffrey D. and {Stello}, Dennis and {Zucker}, Daniel B. and {Zwitter}, Toma{\v{z}} and {Beeson}, Kevin L. and {Buck}, Tobias and {Casagrande}, Luca and {Clark}, Jake T. and {{\v{C}}otar}, Klemen and {da Costa}, Gary S. and {de Grijs}, Richard and {Feuillet}, Diane and {Horner}, Jonathan and {Kafle}, Prajwal R. and {Khanna}, Shourya and {Kobayashi}, Chiaki and {Liu}, Fan and {Montet}, Benjamin T. and {Nandakumar}, Govind and {Nataf}, David M. and {Ness}, Melissa K. and {Spina}, Lorenzo and {Tepper-Garc{\'\i}a}, Thor and {Ting}, Yuan-Sen and {Traven}, Gregor and {Vogrin{\v{c}}i{\v{c}}}, Rok and {Wittenmyer}, Robert A. and {Wyse}, Rosemary F.~G. and {{\v{Z}}erjal}, Maru{\v{s}}a and {Galah Collaboration}},
        title = "{The GALAH+ survey: Third data release}",
      journal = {\mnras},
         year = 2021,
        month = sep,
       volume = {506},
       number = {1},
        pages = {150-201},
          doi = {10.1093/mnras/stab1242},
archivePrefix = {arXiv},
       eprint = {2011.02505},
 primaryClass = {astro-ph.GA},
       adsurl = {https://ui.adsabs.harvard.edu/abs/2021MNRAS.506..150B}
}

@ARTICLE{Camarota2014,
       author = {{Camarota}, L. and {Holberg}, J.~B.},
        title = "{White-dwarf-based evaluation of the GALEX absolute calibration}",
      journal = {\mnras},
         year = 2014,
        month = mar,
       volume = {438},
       number = {4},
        pages = {3111-3118},
          doi = {10.1093/mnras/stt2422},
archivePrefix = {arXiv},
       eprint = {1312.3882},
 primaryClass = {astro-ph.IM},
       adsurl = {https://ui.adsabs.harvard.edu/abs/2014MNRAS.438.3111C}
}

@ARTICLE{Caffau2020,
       author = {{Caffau}, E. and {Bonifacio}, P. and {Sbordone}, L. and {Matas Pinto}, A.~M. and {Fran{\c{c}}ois}, P. and {Jablonka}, P. and {Lardo}, C. and {Martin}, N.~F. and {Starkenburg}, E. and {Aguado}, D. and {Gonz{\'a}lez-Hern{\'a}ndez}, J.~I. and {Venn}, K. and {Mashonkina}, L. and {Sestito}, F.},
        title = "{The Pristine survey XI: the FORS2 sample}",
      journal = {\mnras},
         year = 2020,
        month = apr,
       volume = {493},
       number = {4},
        pages = {4677-4691},
          doi = {10.1093/mnras/staa589},
       adsurl = {https://ui.adsabs.harvard.edu/abs/2020MNRAS.493.4677C}
}

@ARTICLE{Ceccarelli2024,
       author = {{Ceccarelli}, E. and {Massari}, D. and {Mucciarelli}, A. and {Bellazzini}, M. and {Nunnari}, A. and {Cusano}, F. and {Lardo}, C. and {Romano}, D. and {Ilyin}, I. and {Stokholm}, A.},
        title = "{A Walk on the Retrograde Side (WRS) project. I. Tidying-up the retrograde halo with high-resolution spectroscopy}",
      journal = {\aap},
         year = 2024,
        month = apr,
       volume = {684},
          eid = {A37},
        pages = {A37},
          doi = {10.1051/0004-6361/202348332},
archivePrefix = {arXiv},
       eprint = {2401.04184},
 primaryClass = {astro-ph.GA},
       adsurl = {https://ui.adsabs.harvard.edu/abs/2024A&A...684A..37C}
}

@ARTICLE{Chen2020,
       author = {{Chen}, Yan-Ping and {Yan}, Renbin and {Maraston}, Claudia and {Thomas}, Daniel and {Stringfellow}, Guy S. and {Bizyaev}, Dmitry and {Gelfand}, Joseph D. and {Beers}, Timothy C. and {Fern{\'a}ndez-Trincado}, Jos{\'e} G. and {Lazarz}, Daniel and {Hill}, Lewis and {Drory}, Niv and {Stassun}, Keivan G.},
        title = "{Stellar Parameters for the First Release of the MaSTar Library: An Empirical Approach}",
      journal = {\apj},
         year = 2020,
        month = aug,
       volume = {899},
       number = {1},
          eid = {62},
        pages = {62},
          doi = {10.3847/1538-4357/ab9f35},
archivePrefix = {arXiv},
       eprint = {2006.13711},
 primaryClass = {astro-ph.GA},
       adsurl = {https://ui.adsabs.harvard.edu/abs/2020ApJ...899...62C}
}

@ARTICLE{Choi2016,
       author = {{Choi}, Jieun and {Dotter}, Aaron and {Conroy}, Charlie and {Cantiello}, Matteo and {Paxton}, Bill and {Johnson}, Benjamin D.},
        title = "{Mesa Isochrones and Stellar Tracks (MIST). I. Solar-scaled Models}",
      journal = {\apj},
         year = 2016,
        month = jun,
       volume = {823},
       number = {2},
          eid = {102},
        pages = {102},
          doi = {10.3847/0004-637X/823/2/102},
archivePrefix = {arXiv},
       eprint = {1604.08592},
 primaryClass = {astro-ph.SR},
       adsurl = {https://ui.adsabs.harvard.edu/abs/2016ApJ...823..102C}
}

@ARTICLE{Dietz2020,
       author = {{Dietz}, Sarah E. and {Yoon}, Jinmi and {Beers}, Timothy C. and {Placco}, Vinicius M.},
        title = "{The Metallicity Gradient and Complex Formation History of the Outermost Halo of the Milky Way}",
      journal = {\apj},
         year = 2020,
        month = may,
       volume = {894},
       number = {1},
          eid = {34},
        pages = {34},
          doi = {10.3847/1538-4357/ab7fa4},
archivePrefix = {arXiv},
       eprint = {1911.11140},
 primaryClass = {astro-ph.GA},
       adsurl = {https://ui.adsabs.harvard.edu/abs/2020ApJ...894...34D}
}

@BOOK{Kurucz1993,
       author = {{Kurucz}, Robert L.},
        title = "{SYNTHE spectrum synthesis programs and line data}",
         year = 1993,
       adsurl = {https://ui.adsabs.harvard.edu/abs/1993sssp.book.....K}
}

@ARTICLE{Beers2005,
       author = {{Beers}, Timothy C. and {Christlieb}, Norbert},
        title = "{The Discovery and Analysis of Very Metal-Poor Stars in the Galaxy}",
      journal = {\araa},
         year = 2005,
        month = sep,
       volume = {43},
       number = {1},
        pages = {531-580},
          doi = {10.1146/annurev.astro.42.053102.134057},
       adsurl = {https://ui.adsabs.harvard.edu/abs/2005ARA&A..43..531B}
}

@ARTICLE{Abohalima2018,
       author = {{Abohalima}, Abdu and {Frebel}, Anna},
        title = "{JINAbase{\textemdash}A Database for Chemical Abundances of Metal-poor Stars}",
      journal = {\apjs},
         year = 2018,
        month = oct,
       volume = {238},
       number = {2},
          eid = {36},
        pages = {36},
          doi = {10.3847/1538-4365/aadfe9},
archivePrefix = {arXiv},
       eprint = {1711.04410},
 primaryClass = {astro-ph.SR},
       adsurl = {https://ui.adsabs.harvard.edu/abs/2018ApJS..238...36A}
}

@ARTICLE{Mohammed2019,
       author = {{Mohammed}, Steven and {Schiminovich}, David and {Hawkins}, Keith and {Johnson}, Benjamin and {Wang}, Dun and {Hogg}, David W.},
        title = "{An Ultraviolet-Optical Color-Metallicity Relation for Red Clump Stars Using GALEX and Gaia}",
      journal = {\apj},
         year = 2019,
        month = feb,
       volume = {872},
       number = {1},
          eid = {95},
        pages = {95},
          doi = {10.3847/1538-4357/aaf236},
archivePrefix = {arXiv},
       eprint = {1805.03236},
 primaryClass = {astro-ph.SR},
       adsurl = {https://ui.adsabs.harvard.edu/abs/2019ApJ...872...95M}
}

@ARTICLE{Rene2023,
       author = {{Andrae}, Ren{\'e} and {Rix}, Hans-Walter and {Chandra}, Vedant},
        title = "{Robust Data-driven Metallicities for 175 Million Stars from Gaia XP Spectra}",
      journal = {\apjs},
         year = 2023,
        month = jul,
       volume = {267},
       number = {1},
          eid = {8},
        pages = {8},
          doi = {10.3847/1538-4365/acd53e},
archivePrefix = {arXiv},
       eprint = {2302.02611},
 primaryClass = {astro-ph.SR},
       adsurl = {https://ui.adsabs.harvard.edu/abs/2023ApJS..267....8A}
}

@ARTICLE{Sreejith2020,
       author = {{Sreejith}, A.~G. and {Fossati}, L. and {Youngblood}, A. and {France}, K. and {Ambily}, S.},
        title = "{Ca II H\&K stellar activity parameter: a proxy for extreme ultraviolet stellar fluxes}",
      journal = {\aap},
         year = 2020,
        month = dec,
       volume = {644},
          eid = {A67},
        pages = {A67},
          doi = {10.1051/0004-6361/202039167},
archivePrefix = {arXiv},
       eprint = {2010.16179},
 primaryClass = {astro-ph.SR},
       adsurl = {https://ui.adsabs.harvard.edu/abs/2020A&A...644A..67S}
}

@ARTICLE{Deland2012,
       author = {{Deland}, Matthew T. and {Cebula}, Richard P.},
        title = "{Solar UV variations during the decline of Cycle 23}",
      journal = {Journal of Atmospheric and Solar-Terrestrial Physics},
         year = 2012,
        month = mar,
       volume = {77},
        pages = {225-234},
          doi = {10.1016/j.jastp.2012.01.007},
       adsurl = {https://ui.adsabs.harvard.edu/abs/2012JASTP..77..225D}
}

@ARTICLE{Sofia1989,
       author = {{Sofia}, Ulysses J. and {Bruhweiler}, Frederick C. and {Sofia}, Sabatino},
        title = "{Ultraviolet variability of the solar analogue star Alpha Centauri A}",
      journal = {\jgr},
         year = 1989,
        month = jul,
       volume = {94},
       number = {A7},
        pages = {9117-9124},
          doi = {10.1029/JA094iA07p09117},
       adsurl = {https://ui.adsabs.harvard.edu/abs/1989JGR....94.9117S}
}

@ARTICLE{deLaverny2012,
       author = {{de Laverny}, P. and {Recio-Blanco}, A. and {Worley}, C.~C. and {Plez}, B.},
        title = "{The AMBRE project: A new synthetic grid of high-resolution FGKM stellar spectra}",
      journal = {\aap},
         year = 2012,
        month = aug,
       volume = {544},
          eid = {A126},
        pages = {A126},
          doi = {10.1051/0004-6361/201219330},
archivePrefix = {arXiv},
       eprint = {1205.2270},
 primaryClass = {astro-ph.SR},
       adsurl = {https://ui.adsabs.harvard.edu/abs/2012A&A...544A.126D}
}

@ARTICLE{Andrae2023,
       author = {{Andrae}, R. and {Fouesneau}, M. and {Sordo}, R. and {Bailer-Jones}, C.~A.~L. and {Dharmawardena}, T.~E. and {Rybizki}, J. and {De Angeli}, F. and {Lindstr{\o}m}, H.~E.~P. and {Marshall}, D.~J. and {Drimmel}, R. and {Korn}, A.~J. and {Soubiran}, C. and {Brouillet}, N. and {Casamiquela}, L. and {Rix}, H.-W. and {Abreu Aramburu}, A. and {{\'A}lvarez}, M.~A. and {Bakker}, J. and {Bellas-Velidis}, I. and {Bijaoui}, A. and {Brugaletta}, E. and {Burlacu}, A. and {Carballo}, R. and {Chaoul}, L. and {Chiavassa}, A. and {Contursi}, G. and {Cooper}, W.~J. and {Creevey}, O.~L. and {Dafonte}, C. and {Dapergolas}, A. and {de Laverny}, P. and {Delchambre}, L. and {Demouchy}, C. and {Edvardsson}, B. and {Fr{\'e}mat}, Y. and {Garabato}, D. and {Garc{\'\i}a-Lario}, P. and {Garc{\'\i}a-Torres}, M. and {Gavel}, A. and {Gomez}, A. and {Gonz{\'a}lez-Santamar{\'\i}a}, I. and {Hatzidimitriou}, D. and {Heiter}, U. and {Jean-Antoine Piccolo}, A. and {Kontizas}, M. and {Kordopatis}, G. and {Lanzafame}, A.~C. and {Lebreton}, Y. and {Licata}, E.~L. and {Livanou}, E. and {Lobel}, A. and {Lorca}, A. and {Magdaleno Romeo}, A. and {Manteiga}, M. and {Marocco}, F. and {Mary}, N. and {Nicolas}, C. and {Ordenovic}, C. and {Pailler}, F. and {Palicio}, P.~A. and {Pallas-Quintela}, L. and {Panem}, C. and {Pichon}, B. and {Poggio}, E. and {Recio-Blanco}, A. and {Riclet}, F. and {Robin}, C. and {Santove{\~n}a}, R. and {Sarro}, L.~M. and {Schultheis}, M.~S. and {Segol}, M. and {Silvelo}, A. and {Slezak}, I. and {Smart}, R.~L. and {S{\"u}veges}, M. and {Th{\'e}venin}, F. and {Torralba Elipe}, G. and {Ulla}, A. and {Utrilla}, E. and {Vallenari}, A. and {van Dillen}, E. and {Zhao}, H. and {Zorec}, J.},
        title = "{Gaia Data Release 3. Analysis of the Gaia BP/RP spectra using the General Stellar Parameterizer from Photometry}",
      journal = {\aap},
         year = 2023,
        month = jun,
       volume = {674},
          eid = {A27},
        pages = {A27},
          doi = {10.1051/0004-6361/202243462},
archivePrefix = {arXiv},
       eprint = {2206.06138},
 primaryClass = {astro-ph.SR},
       adsurl = {https://ui.adsabs.harvard.edu/abs/2023A&A...674A..27A}
}

@ARTICLE{Gaia2023,
       author = {{Gaia Collaboration} and {Vallenari}, A. and {Brown}, A.~G.~A. and {Prusti}, T. and {de Bruijne}, J.~H.~J. and {Arenou}, F. and {Babusiaux}, C. and {Biermann}, M. and {Creevey}, O.~L. and {Ducourant}, C. and {Evans}, D.~W. and {Eyer}, L. and {Guerra}, R. and {Hutton}, A. and {Jordi}, C. and {Klioner}, S.~A. and {Lammers}, U.~L. and {Lindegren}, L. and {Luri}, X. and {Mignard}, F. and {Panem}, C. and {Pourbaix}, D. and {Randich}, S. and {Sartoretti}, P. and {Soubiran}, C. and {Tanga}, P. and {Walton}, N.~A. and {Bailer-Jones}, C.~A.~L. and {Bastian}, U. and {Drimmel}, R. and {Jansen}, F. and {Katz}, D. and {Lattanzi}, M.~G. and {van Leeuwen}, F. and {Bakker}, J. and {Cacciari}, C. and {Casta{\~n}eda}, J. and {De Angeli}, F. and {Fabricius}, C. and {Fouesneau}, M. and {Fr{\'e}mat}, Y. and {Galluccio}, L. and {Guerrier}, A. and {Heiter}, U. and {Masana}, E. and {Messineo}, R. and {Mowlavi}, N. and {Nicolas}, C. and {Nienartowicz}, K. and {Pailler}, F. and {Panuzzo}, P. and {Riclet}, F. and {Roux}, W. and {Seabroke}, G.~M. and {Sordo}, R. and {Th{\'e}venin}, F. and {Gracia-Abril}, G. and {Portell}, J. and {Teyssier}, D. and {Altmann}, M. and {Andrae}, R. and {Audard}, M. and {Bellas-Velidis}, I. and {Benson}, K. and {Berthier}, J. and {Blomme}, R. and {Burgess}, P.~W. and {Busonero}, D. and {Busso}, G. and {C{\'a}novas}, H. and {Carry}, B. and {Cellino}, A. and {Cheek}, N. and {Clementini}, G. and {Damerdji}, Y. and {Davidson}, M. and {de Teodoro}, P. and {Nu{\~n}ez Campos}, M. and {Delchambre}, L. and {Dell'Oro}, A. and {Esquej}, P. and {Fern{\'a}ndez-Hern{\'a}ndez}, J. and {Fraile}, E. and {Garabato}, D. and {Garc{\'\i}a-Lario}, P. and {Gosset}, E. and {Haigron}, R. and {Halbwachs}, J. -L. and {Hambly}, N.~C. and {Harrison}, D.~L. and {Hern{\'a}ndez}, J. and {Hestroffer}, D. and {Hodgkin}, S.~T. and {Holl}, B. and {Jan{\ss}en}, K. and {Jevardat de Fombelle}, G. and {Jordan}, S. and {Krone-Martins}, A. and {Lanzafame}, A.~C. and {L{\"o}ffler}, W. and {Marchal}, O. and {Marrese}, P.~M. and {Moitinho}, A. and {Muinonen}, K. and {Osborne}, P. and {Pancino}, E. and {Pauwels}, T. and {Recio-Blanco}, A. and {Reyl{\'e}}, C. and {Riello}, M. and {Rimoldini}, L. and {Roegiers}, T. and {Rybizki}, J. and {Sarro}, L.~M. and {Siopis}, C. and {Smith}, M. and {Sozzetti}, A. and {Utrilla}, E. and {van Leeuwen}, M. and {Abbas}, U. and {{\'A}brah{\'a}m}, P. and {Abreu Aramburu}, A. and {Aerts}, C. and {Aguado}, J.~J. and {Ajaj}, M. and {Aldea-Montero}, F. and {Altavilla}, G. and {{\'A}lvarez}, M.~A. and {Alves}, J. and {Anders}, F. and {Anderson}, R.~I. and {Anglada Varela}, E. and {Antoja}, T. and {Baines}, D. and {Baker}, S.~G. and {Balaguer-N{\'u}{\~n}ez}, L. and {Balbinot}, E. and {Balog}, Z. and {Barache}, C. and {Barbato}, D. and {Barros}, M. and {Barstow}, M.~A. and {Bartolom{\'e}}, S. and {Bassilana}, J. -L. and {Bauchet}, N. and {Becciani}, U. and {Bellazzini}, M. and {Berihuete}, A. and {Bernet}, M. and {Bertone}, S. and {Bianchi}, L. and {Binnenfeld}, A. and {Blanco-Cuaresma}, S. and {Blazere}, A. and {Boch}, T. and {Bombrun}, A. and {Bossini}, D. and {Bouquillon}, S. and {Bragaglia}, A. and {Bramante}, L. and {Breedt}, E. and {Bressan}, A. and {Brouillet}, N. and {Brugaletta}, E. and {Bucciarelli}, B. and {Burlacu}, A. and {Butkevich}, A.~G. and {Buzzi}, R. and {Caffau}, E. and {Cancelliere}, R. and {Cantat-Gaudin}, T. and {Carballo}, R. and {Carlucci}, T. and {Carnerero}, M.~I. and {Carrasco}, J.~M. and {Casamiquela}, L. and {Castellani}, M. and {Castro-Ginard}, A. and {Chaoul}, L. and {Charlot}, P. and {Chemin}, L. and {Chiaramida}, V. and {Chiavassa}, A. and {Chornay}, N. and {Comoretto}, G. and {Contursi}, G. and {Cooper}, W.~J. and {Cornez}, T. and {Cowell}, S. and {Crifo}, F. and {Cropper}, M. and {Crosta}, M. and {Crowley}, C. and {Dafonte}, C. and {Dapergolas}, A. and {David}, M. and {David}, P. and {de Laverny}, P. and {De Luise}, F. and {De March}, R.},
        title = "{Gaia Data Release 3. Summary of the content and survey properties}",
      journal = {\aap},
         year = 2023,
        month = jun,
       volume = {674},
          eid = {A1},
        pages = {A1},
          doi = {10.1051/0004-6361/202243940},
archivePrefix = {arXiv},
       eprint = {2208.00211},
 primaryClass = {astro-ph.GA},
       adsurl = {https://ui.adsabs.harvard.edu/abs/2023A&A...674A...1G}
}

@ARTICLE{Green2019,
       author = {{Green}, Gregory M. and {Schlafly}, Edward and {Zucker}, Catherine and {Speagle}, Joshua S. and {Finkbeiner}, Douglas},
        title = "{A 3D Dust Map Based on Gaia, Pan-STARRS 1, and 2MASS}",
      journal = {\apj},
         year = 2019,
        month = dec,
       volume = {887},
       number = {1},
          eid = {93},
        pages = {93},
          doi = {10.3847/1538-4357/ab5362},
archivePrefix = {arXiv},
       eprint = {1905.02734},
 primaryClass = {astro-ph.GA},
       adsurl = {https://ui.adsabs.harvard.edu/abs/2019ApJ...887...93G}
}

@ARTICLE{Gu2015,
       author = {{Gu}, Jiayin and {Du}, Cuihua and {Jia}, Yunpeng and {Peng}, Xiyan and {Wu}, Zhenyu and {Jing}, Yingjie and {Ma}, Jun and {Zhou}, Xu and {Fan}, Xiaohui and {Fan}, Zhou and {Jing}, Yipeng and {Jiang}, Zhaoji and {Lesser}, Michael and {Nie}, Jundan and {Shen}, Shiyin and {Wang}, Jiali and {Zou}, Hu and {Zhang}, Tianmeng and {Zhou}, Zhimin},
        title = "{Photometric metallicity calibration with SDSS and SCUSS and its application to distant stars in the south Galactic cap}",
      journal = {\mnras},
         year = 2015,
        month = sep,
       volume = {452},
       number = {3},
        pages = {3092-3099},
          doi = {10.1093/mnras/stv1529},
archivePrefix = {arXiv},
       eprint = {1507.02054},
 primaryClass = {astro-ph.GA},
       adsurl = {https://ui.adsabs.harvard.edu/abs/2015MNRAS.452.3092G}
}

@ARTICLE{Husser2013,
       author = {{Husser}, T. -O. and {Wende-von Berg}, S. and {Dreizler}, S. and {Homeier}, D. and {Reiners}, A. and {Barman}, T. and {Hauschildt}, P.~H.},
        title = "{A new extensive library of PHOENIX stellar atmospheres and synthetic spectra}",
      journal = {\aap},
         year = 2013,
        month = may,
       volume = {553},
          eid = {A6},
        pages = {A6},
          doi = {10.1051/0004-6361/201219058},
archivePrefix = {arXiv},
       eprint = {1303.5632},
 primaryClass = {astro-ph.SR},
       adsurl = {https://ui.adsabs.harvard.edu/abs/2013A&A...553A...6H}
}

@ARTICLE{Hourihane2023,
       author = {{Hourihane}, A. and {Fran{\c{c}}ois}, P. and {Worley}, C.~C. and {Magrini}, L. and {Gonneau}, A. and {Casey}, A.~R. and {Gilmore}, G. and {Randich}, S. and {Sacco}, G.~G. and {Recio-Blanco}, A. and {Korn}, A.~J. and {Allende Prieto}, C. and {Smiljanic}, R. and {Blomme}, R. and {Bragaglia}, A. and {Walton}, N.~A. and {Van Eck}, S. and {Bensby}, T. and {Lanzafame}, A. and {Frasca}, A. and {Franciosini}, E. and {Damiani}, F. and {Lind}, K. and {Bergemann}, M. and {Bonifacio}, P. and {Hill}, V. and {Lobel}, A. and {Montes}, D. and {Feuillet}, D.~K. and {Tautvai{\v{s}}ien{\.{e}}}, G. and {Guiglion}, G. and {Tabernero}, H.~M. and {Gonz{\'a}lez Hern{\'a}ndez}, J.~I. and {Gebran}, M. and {Van der Swaelmen}, M. and {Mikolaitis}, {\v{S}}. and {Daflon}, S. and {Merle}, T. and {Morel}, T. and {Lewis}, J.~R. and {Gonz{\'a}lez Solares}, E.~A. and {Murphy}, D.~N.~A. and {Jeffries}, R.~D. and {Jackson}, R.~J. and {Feltzing}, S. and {Prusti}, T. and {Carraro}, G. and {Biazzo}, K. and {Prisinzano}, L. and {Jofr{\'e}}, P. and {Zaggia}, S. and {Drazdauskas}, A. and {Stonkut{\'e}}, E. and {Marfil}, E. and {Jim{\'e}nez-Esteban}, F. and {Mahy}, L. and {Guti{\'e}rrez Albarr{\'a}n}, M.~L. and {Berlanas}, S.~R. and {Santos}, W. and {Morbidelli}, L. and {Spina}, L. and {Minkevi{\v{c}}i{\={u}}t{\.{e}}}, R.},
        title = "{The Gaia-ESO Survey: Homogenisation of stellar parameters and elemental abundances}",
      journal = {\aap},
         year = 2023,
        month = aug,
       volume = {676},
          eid = {A129},
        pages = {A129},
          doi = {10.1051/0004-6361/202345910},
archivePrefix = {arXiv},
       eprint = {2304.07720},
 primaryClass = {astro-ph.SR},
       adsurl = {https://ui.adsabs.harvard.edu/abs/2023A&A...676A.129H}
}

@ARTICLE{Ishigaki2013,
       author = {{Ishigaki}, M.~N. and {Aoki}, W. and {Chiba}, M.},
        title = "{Chemical Abundances of the Milky Way Thick Disk and Stellar Halo. II. Sodium, Iron-peak, and Neutron-capture Elements}",
      journal = {\apj},
         year = 2013,
        month = jul,
       volume = {771},
       number = {1},
          eid = {67},
        pages = {67},
          doi = {10.1088/0004-637X/771/1/67},
archivePrefix = {arXiv},
       eprint = {1306.0954},
 primaryClass = {astro-ph.GA},
       adsurl = {https://ui.adsabs.harvard.edu/abs/2013ApJ...771...67I}
}

@ARTICLE{Ivezic2008,
       author = {{Ivezi{\'c}}, {\v{Z}}eljko and {Sesar}, Branimir and {Juri{\'c}}, Mario and {Bond}, Nicholas and {Dalcanton}, Julianne and {Rockosi}, Constance M. and {Yanny}, Brian and {Newberg}, Heidi J. and {Beers}, Timothy C. and {Allende Prieto}, Carlos and {Wilhelm}, Ron and {Lee}, Young Sun and {Sivarani}, Thirupathi and {Norris}, John E. and {Bailer-Jones}, Coryn A.~L. and {Re Fiorentin}, Paola and {Schlegel}, David and {Uomoto}, Alan and {Lupton}, Robert H. and {Knapp}, Gillian R. and {Gunn}, James E. and {Covey}, Kevin R. and {Allyn Smith}, J. and {Miknaitis}, Gajus and {Doi}, Mamoru and {Tanaka}, Masayuki and {Fukugita}, Masataka and {Kent}, Steve and {Finkbeiner}, Douglas and {Munn}, Jeffrey A. and {Pier}, Jeffrey R. and {Quinn}, Tom and {Hawley}, Suzanne and {Anderson}, Scott and {Kiuchi}, Furea and {Chen}, Alex and {Bushong}, James and {Sohi}, Harkirat and {Haggard}, Daryl and {Kimball}, Amy and {Barentine}, John and {Brewington}, Howard and {Harvanek}, Mike and {Kleinman}, Scott and {Krzesinski}, Jurek and {Long}, Dan and {Nitta}, Atsuko and {Snedden}, Stephanie and {Lee}, Brian and {Harris}, Hugh and {Brinkmann}, Jonathan and {Schneider}, Donald P. and {York}, Donald G.},
        title = "{The Milky Way Tomography with SDSS. II. Stellar Metallicity}",
      journal = {\apj},
         year = 2008,
        month = sep,
       volume = {684},
       number = {1},
        pages = {287-325},
          doi = {10.1086/589678},
archivePrefix = {arXiv},
       eprint = {0804.3850},
 primaryClass = {astro-ph},
       adsurl = {https://ui.adsabs.harvard.edu/abs/2008ApJ...684..287I}
}

@ARTICLE{Jonsson2020,
       author = {{J{\"o}nsson}, Henrik and {Holtzman}, Jon A. and {Allende Prieto}, Carlos and {Cunha}, Katia and {Garc{\'\i}a-Hern{\'a}ndez}, D.~A. and {Hasselquist}, Sten and {Masseron}, Thomas and {Osorio}, Yeisson and {Shetrone}, Matthew and {Smith}, Verne and {Stringfellow}, Guy S. and {Bizyaev}, Dmitry and {Edvardsson}, Bengt and {Majewski}, Steven R. and {M{\'e}sz{\'a}ros}, Szabolcs and {Souto}, Diogo and {Zamora}, Olga and {Beaton}, Rachael L. and {Bovy}, Jo and {Donor}, John and {Pinsonneault}, Marc H. and {Poovelil}, Vijith Jacob and {Sobeck}, Jennifer},
        title = "{APOGEE Data and Spectral Analysis from SDSS Data Release 16: Seven Years of Observations Including First Results from APOGEE-South}",
      journal = {\aj},
         year = 2020,
        month = sep,
       volume = {160},
       number = {3},
          eid = {120},
        pages = {120},
          doi = {10.3847/1538-3881/aba592},
archivePrefix = {arXiv},
       eprint = {2007.05537},
 primaryClass = {astro-ph.GA},
       adsurl = {https://ui.adsabs.harvard.edu/abs/2020AJ....160..120J}
}

@ARTICLE{Lai2008,
       author = {{Lai}, David K. and {Bolte}, Michael and {Johnson}, Jennifer A. and {Lucatello}, Sara and {Heger}, Alexander and {Woosley}, S.~E.},
        title = "{Detailed Abundances for 28 Metal-poor Stars: Stellar Relics in the Milky Way}",
      journal = {\apj},
         year = 2008,
        month = jul,
       volume = {681},
       number = {2},
        pages = {1524-1556},
          doi = {10.1086/588811},
archivePrefix = {arXiv},
       eprint = {0804.1370},
 primaryClass = {astro-ph},
       adsurl = {https://ui.adsabs.harvard.edu/abs/2008ApJ...681.1524L}
}

@ARTICLE{Li2022,
       author = {{Li}, Haining and {Aoki}, Wako and {Matsuno}, Tadafumi and {Xing}, Qianfan and {Suda}, Takuma and {Tominaga}, Nozomu and {Chen}, Yuqin and {Honda}, Satoshi and {Ishigaki}, Miho N. and {Shi}, Jianrong and {Zhao}, Jingkun and {Zhao}, Gang},
        title = "{Four-hundred Very Metal-poor Stars Studied with LAMOST and Subaru. II. Elemental Abundances}",
      journal = {\apj},
         year = 2022,
        month = jun,
       volume = {931},
       number = {2},
          eid = {147},
        pages = {147},
          doi = {10.3847/1538-4357/ac6514},
archivePrefix = {arXiv},
       eprint = {2203.11529},
 primaryClass = {astro-ph.SR},
       adsurl = {https://ui.adsabs.harvard.edu/abs/2022ApJ...931..147L}
}

@ARTICLE{Limberg2021,
       author = {{Limberg}, Guilherme and {Rossi}, Silvia and {Beers}, Timothy C. and {Perottoni}, H{\'e}lio D. and {P{\'e}rez-Villegas}, Angeles and {Santucci}, Rafael M. and {Abuchaim}, Yuri and {Placco}, Vinicius M. and {Lee}, Young Sun and {Christlieb}, Norbert and {Norris}, John E. and {Bessell}, Michael S. and {Ryan}, Sean G. and {Wilhelm}, Ronald and {Rhee}, Jaehyon and {Frebel}, Anna},
        title = "{Dynamically Tagged Groups of Very Metal-poor Halo Stars from the HK and Hamburg/ESO Surveys}",
      journal = {\apj},
         year = 2021,
        month = jan,
       volume = {907},
       number = {1},
          eid = {10},
        pages = {10},
          doi = {10.3847/1538-4357/abcb87},
archivePrefix = {arXiv},
       eprint = {2011.08305},
 primaryClass = {astro-ph.GA},
       adsurl = {https://ui.adsabs.harvard.edu/abs/2021ApJ...907...10L}
}

@ARTICLE{Lucchesi2022,
       author = {{Lucchesi}, R. and {Lardo}, C. and {Jablonka}, P. and {Sestito}, F. and {Mashonkina}, L. and {Arentsen}, A. and {Suter}, W. and {Venn}, K. and {Martin}, N. and {Starkenburg}, E. and {Aguado}, D. and {Hill}, V. and {Kordopatis}, G. and {Navarro}, J.~F. and {Gonz{\'a}lez Hern{\'a}ndez}, J.~I. and {Malhan}, K. and {Yuan}, Z.},
        title = "{The Pristine survey - XV. A CFHT ESPaDOnS view on the Milky Way halo and disc populations}",
      journal = {\mnras},
         year = 2022,
        month = mar,
       volume = {511},
       number = {1},
        pages = {1004-1021},
          doi = {10.1093/mnras/stab3721},
archivePrefix = {arXiv},
       eprint = {2112.10792},
 primaryClass = {astro-ph.GA},
       adsurl = {https://ui.adsabs.harvard.edu/abs/2022MNRAS.511.1004L}
}

@ARTICLE{Mardini2024,
       author = {{Mardini}, Mohammad K. and {Frebel}, Anna and {Betre}, Leyatt and {Jacobson}, Heather and {Norris}, John E. and {Christlieb}, Norbert},
        title = "{Metal-poor stars observed with the Magellan Telescope - IV. Neutron-capture element signatures in 27 main-sequence stars}",
      journal = {\mnras},
         year = 2024,
        month = feb,
       volume = {528},
       number = {2},
        pages = {2912-2929},
          doi = {10.1093/mnras/stad3925},
archivePrefix = {arXiv},
       eprint = {2305.05363},
 primaryClass = {astro-ph.GA},
       adsurl = {https://ui.adsabs.harvard.edu/abs/2024MNRAS.528.2912M}
}

@ARTICLE{Martin2005,
       author = {{Martin}, D. Christopher and {Fanson}, James and {Schiminovich}, David and {Morrissey}, Patrick and {Friedman}, Peter G. and {Barlow}, Tom A. and {Conrow}, Tim and {Grange}, Robert and {Jelinsky}, Patrick N. and {Milliard}, Bruno and {Siegmund}, Oswald H.~W. and {Bianchi}, Luciana and {Byun}, Yong-Ik and {Donas}, Jose and {Forster}, Karl and {Heckman}, Timothy M. and {Lee}, Young-Wook and {Madore}, Barry F. and {Malina}, Roger F. and {Neff}, Susan G. and {Rich}, R. Michael and {Small}, Todd and {Surber}, Frank and {Szalay}, Alex S. and {Welsh}, Barry and {Wyder}, Ted K.},
        title = "{The Galaxy Evolution Explorer: A Space Ultraviolet Survey Mission}",
      journal = {\apjl},
         year = 2005,
        month = jan,
       volume = {619},
       number = {1},
        pages = {L1-L6},
          doi = {10.1086/426387},
archivePrefix = {arXiv},
       eprint = {astro-ph/0411302},
 primaryClass = {astro-ph},
       adsurl = {https://ui.adsabs.harvard.edu/abs/2005ApJ...619L...1M}
}

@ARTICLE{Matteucci2021,
       author = {{Matteucci}, Francesca},
        title = "{Modelling the chemical evolution of the Milky Way}",
      journal = {\aapr},
         year = 2021,
        month = dec,
       volume = {29},
       number = {1},
          eid = {5},
        pages = {5},
          doi = {10.1007/s00159-021-00133-8},
archivePrefix = {arXiv},
       eprint = {2106.13145},
 primaryClass = {astro-ph.GA},
       adsurl = {https://ui.adsabs.harvard.edu/abs/2021A&ARv..29....5M}
}

@ARTICLE{Moe2019,
       author = {{Moe}, Maxwell and {Kratter}, Kaitlin M. and {Badenes}, Carles},
        title = "{The Close Binary Fraction of Solar-type Stars Is Strongly Anticorrelated with Metallicity}",
      journal = {\apj},
         year = 2019,
        month = apr,
       volume = {875},
       number = {1},
          eid = {61},
        pages = {61},
          doi = {10.3847/1538-4357/ab0d88},
archivePrefix = {arXiv},
       eprint = {1808.02116},
 primaryClass = {astro-ph.SR},
       adsurl = {https://ui.adsabs.harvard.edu/abs/2019ApJ...875...61M}
}

@ARTICLE{Palacios2010,
       author = {{Palacios}, A. and {Gebran}, M. and {Josselin}, E. and {Martins}, F. and {Plez}, B. and {Belmas}, M. and {L{\`e}bre}, A.},
        title = "{POLLUX: a database of synthetic stellar spectra}",
      journal = {\aap},
         year = 2010,
        month = jun,
       volume = {516},
          eid = {A13},
        pages = {A13},
          doi = {10.1051/0004-6361/200913932},
archivePrefix = {arXiv},
       eprint = {1003.4682},
 primaryClass = {astro-ph.SR},
       adsurl = {https://ui.adsabs.harvard.edu/abs/2010A&A...516A..13P}
}

@ARTICLE{PecautMamajek2013,
       author = {{Pecaut}, Mark J. and {Mamajek}, Eric E.},
        title = "{Intrinsic Colors, Temperatures, and Bolometric Corrections of Pre-main-sequence Stars}",
      journal = {\apjs},
         year = 2013,
        month = sep,
       volume = {208},
       number = {1},
          eid = {9},
        pages = {9},
          doi = {10.1088/0067-0049/208/1/9},
archivePrefix = {arXiv},
       eprint = {1307.2657},
 primaryClass = {astro-ph.SR},
       adsurl = {https://ui.adsabs.harvard.edu/abs/2013ApJS..208....9P}
}

@ARTICLE{Placco2022,
       author = {{Placco}, Vinicius M. and {Almeida-Fernandes}, Felipe and {Arentsen}, Anke and {Lee}, Young Sun and {Schoenell}, William and {Ribeiro}, Tiago and {Kanaan}, Antonio},
        title = "{Mining S-PLUS for Metal-poor Stars in the Milky Way}",
      journal = {\apjs},
         year = 2022,
        month = sep,
       volume = {262},
       number = {1},
          eid = {8},
        pages = {8},
          doi = {10.3847/1538-4365/ac7ab0},
archivePrefix = {arXiv},
       eprint = {2206.09003},
 primaryClass = {astro-ph.SR},
       adsurl = {https://ui.adsabs.harvard.edu/abs/2022ApJS..262....8P}
}

@ARTICLE{Placco2010,
       author = {{Placco}, Vinicius M. and {Kennedy}, Catherine R. and {Rossi}, Silvia and {Beers}, Timothy C. and {Lee}, Young Sun and {Christlieb}, Norbert and {Sivarani}, Thirupathi and {Reimers}, Dieter and {Wisotzki}, Lutz},
        title = "{A Search for Unrecognized Carbon-Enhanced Metal-Poor Stars in the Galaxy}",
      journal = {\aj},
         year = 2010,
        month = mar,
       volume = {139},
       number = {3},
        pages = {1051-1065},
          doi = {10.1088/0004-6256/139/3/1051},
archivePrefix = {arXiv},
       eprint = {1001.2512},
 primaryClass = {astro-ph.SR},
       adsurl = {https://ui.adsabs.harvard.edu/abs/2010AJ....139.1051P}
}

@ARTICLE{Roederer2014,
       author = {{Roederer}, Ian U. and {Preston}, George W. and {Thompson}, Ian B. and {Shectman}, Stephen A. and {Sneden}, Christopher and {Burley}, Gregory S. and {Kelson}, Daniel D.},
        title = "{A Search for Stars of Very Low Metal Abundance. VI. Detailed Abundances of 313 Metal-poor Stars}",
      journal = {\aj},
         year = 2014,
        month = jun,
       volume = {147},
       number = {6},
          eid = {136},
        pages = {136},
          doi = {10.1088/0004-6256/147/6/136},
archivePrefix = {arXiv},
       eprint = {1403.6853},
 primaryClass = {astro-ph.SR},
       adsurl = {https://ui.adsabs.harvard.edu/abs/2014AJ....147..136R}
}

@ARTICLE{Ryan1991,
       author = {{Ryan}, Sean G. and {Norris}, John E.},
        title = "{Subdwarf Studies. II. Abundances and Kinematics from Medium Resolution Spectra}",
      journal = {\aj},
         year = 1991,
        month = may,
       volume = {101},
        pages = {1835},
          doi = {10.1086/115811},
       adsurl = {https://ui.adsabs.harvard.edu/abs/1991AJ....101.1835R}
}

@ARTICLE{Sestito2020,
       author = {{Sestito}, Federico and {Martin}, Nicolas F. and {Starkenburg}, Else and {Arentsen}, Anke and {Ibata}, Rodrigo A. and {Longeard}, Nicolas and {Kielty}, Collin and {Youakim}, Kristopher and {Venn}, Kim A. and {Aguado}, David S. and {Carlberg}, Raymond G. and {Gonz{\'a}lez Hern{\'a}ndez}, Jonay I. and {Hill}, Vanessa and {Jablonka}, Pascale and {Kordopatis}, Georges and {Malhan}, Khyati and {Navarro}, Julio F. and {S{\'a}nchez-Janssen}, Rub{\'e}n and {Thomas}, Guillame and {Tolstoy}, Eline and {Wilson}, Thomas G. and {Palicio}, Pedro A. and {Bialek}, Spencer and {Garcia-Dias}, Rafael and {Lucchesi}, Romain and {North}, Pierre and {Osorio}, Yeisson and {Patrick}, Lee R. and {Peralta de Arriba}, Luis},
        title = "{The Pristine survey - X. A large population of low-metallicity stars permeates the Galactic disc}",
      journal = {\mnras},
         year = 2020,
        month = sep,
       volume = {497},
       number = {1},
        pages = {L7-L12},
          doi = {10.1093/mnrasl/slaa022},
archivePrefix = {arXiv},
       eprint = {1911.08491},
 primaryClass = {astro-ph.GA},
       adsurl = {https://ui.adsabs.harvard.edu/abs/2020MNRAS.497L...7S}
}

@ARTICLE{Shank2022,
       author = {{Shank}, Derek and {Komater}, Dante and {Beers}, Timothy C. and {Placco}, Vinicius M. and {Huang}, Yang},
        title = "{Dynamically Tagged Groups of Metal-poor Stars. II. The Radial Velocity Experiment Data Release 6}",
      journal = {\apjs},
         year = 2022,
        month = aug,
       volume = {261},
       number = {2},
          eid = {19},
        pages = {19},
          doi = {10.3847/1538-4365/ac680c},
archivePrefix = {arXiv},
       eprint = {2201.08337},
 primaryClass = {astro-ph.GA},
       adsurl = {https://ui.adsabs.harvard.edu/abs/2022ApJS..261...19S}
}

@ARTICLE{Shen2023,
       author = {{Shen}, Yu-Fu and {Alexeeva}, S.~A. and {Zhao}, Gang and {Liu}, Shuai and {Zhou}, Zeming and {Yan}, Hongliang and {Li}, Haining and {Chen}, Tianyi and {Xu}, Xiaodong and {Chen}, Huiling and {Zhang}, Huawei and {Shi}, Jianrong},
        title = "{Sodium Abundances in Very Metal-poor Stars}",
      journal = {Research in Astronomy and Astrophysics},
         year = 2023,
        month = jul,
       volume = {23},
       number = {7},
          eid = {075019},
        pages = {075019},
          doi = {10.1088/1674-4527/accdc3},
       adsurl = {https://ui.adsabs.harvard.edu/abs/2023RAA....23g5019S}
}

@ARTICLE{Short2005,
       author = {{Short}, C.~I. and {Hauschildt}, P.~H.},
        title = "{A Non-LTE Line-Blanketed Model of a Solar-Type Star}",
      journal = {\apj},
         year = 2005,
        month = jan,
       volume = {618},
       number = {2},
        pages = {926-938},
          doi = {10.1086/426128},
archivePrefix = {arXiv},
       eprint = {astro-ph/0409693},
 primaryClass = {astro-ph},
       adsurl = {https://ui.adsabs.harvard.edu/abs/2005ApJ...618..926S}
}

@ARTICLE{Schlafly2011,
       author = {{Schlafly}, Edward F. and {Finkbeiner}, Douglas P.},
        title = "{Measuring Reddening with Sloan Digital Sky Survey Stellar Spectra and Recalibrating SFD}",
      journal = {\apj},
         year = 2011,
        month = aug,
       volume = {737},
       number = {2},
          eid = {103},
        pages = {103},
          doi = {10.1088/0004-637X/737/2/103},
archivePrefix = {arXiv},
       eprint = {1012.4804},
 primaryClass = {astro-ph.GA},
       adsurl = {https://ui.adsabs.harvard.edu/abs/2011ApJ...737..103S}
}

@ARTICLE{Soubiran2022,
       author = {{Soubiran}, C. and {Brouillet}, N. and {Casamiquela}, L.},
        title = "{Assessment of [Fe/H] determinations for FGK stars in spectroscopic surveys}",
      journal = {\aap},
         year = 2022,
        month = jul,
       volume = {663},
          eid = {A4},
        pages = {A4},
          doi = {10.1051/0004-6361/202142409},
archivePrefix = {arXiv},
       eprint = {2112.07545},
 primaryClass = {astro-ph.SR},
       adsurl = {https://ui.adsabs.harvard.edu/abs/2022A&A...663A...4S}
}

@ARTICLE{Soubiran2016,
       author = {{Soubiran}, Caroline and {Le Campion}, Jean-Fran{\c{c}}ois and {Brouillet}, Nathalie and {Chemin}, Laurent},
        title = "{The PASTEL catalogue: 2016 version}",
      journal = {\aap},
         year = 2016,
        month = jun,
       volume = {591},
          eid = {A118},
        pages = {A118},
          doi = {10.1051/0004-6361/201628497},
archivePrefix = {arXiv},
       eprint = {1605.07384},
 primaryClass = {astro-ph.SR},
       adsurl = {https://ui.adsabs.harvard.edu/abs/2016A&A...591A.118S}
}

\end{document}